\documentclass[useAMS,onecolumn,usenatbib]{mn2e}

\newif\ifAMStwofonts

\usepackage{graphicx}	
\usepackage{amssymb}	
\usepackage{epsfig}
\usepackage{color}

\def\pmb#1{\mbox{\boldmath$#1$}}
\def\gtsim {>\kern-1.2em\lower1.1ex\hbox{$\sim$}}
\def\ltsim {<\kern-1.2em\lower1.1ex\hbox{$\sim$}}
\def\gtsim {>\kern-1.2em\lower1.1ex\hbox{$\sim$}}
\def\ltsim {<\kern-1.2em\lower1.1ex\hbox{$\sim$}}

\def\be{\begin{equation}}
\def\ee{\end{equation}}

\def\pmbmt#1{\pmb{\sf #1}}
\def\rmi{{\rm i}}

\begin{document}

\title[tidally driven mean flows]{Tidally driven mean flows in slowly and uniformly rotating massive main sequence stars}

\author[U. Lee]{
Umin Lee$^{1}$\thanks{E-mail: lee@astr.tohoku.ac.jp},
\\
$^{1}$Astronomical Institute, Tohoku University, Sendai, Miyagi 980-8578, Japan\\
}

\date{Accepted XXX. Received YYY; in original form ZZZ}
\pubyear{2015}

\maketitle

\begin{abstract}
We calculate tidally driven mean flows in a slowly and uniformly rotating massive main sequence star in a binary system.
We treat the tidal potential due to the companion as a small perturbation to the primary star.
We compute tidal responses of the primary as forced linear oscillations, as a function
of the tidal forcing frequency $\omega_{\rm tide}=2(\Omega_{\rm orb}-\Omega)$, where $\Omega_{\rm orb}$ is
the mean orbital angular velocity and $\Omega$ is the angular velocity of rotation of the primary star.
The amplitude of the tidal responses is proportional to the parameter $f_0\propto (M_2/M)(a_{\rm orb}/R)^{-3}$,
where $M$ and $M_2$ are the masses of the primary and companion stars, $R$ is the radius of the primary and $a_{\rm orb}$
is the mean orbital separation between the stars.
For a given $f_0$, the amplitudes depend on $\omega_{\rm tide}$ and become large when $\omega_{\rm tide}$ is in 
resonance with natural frequencies of the star.
Using the tidal responses, we calculate axisymmetric mean flows, assuming that the mean flows are non-oscillatory flows driven via non-linear effects of linear tidal responses.
We find that the $\phi$-component of the mean flow velocity dominates.
We also find that the amplitudes of the mean flows are large only in the surface layers 
where non-adiabatic effects are significant and that the amplitudes are confined to the equatorial regions of the star.
Depending on $M_2/M$ and $a_{\rm orb}/R$, the amplitudes of mean flows at the surface become significant.

\end{abstract}



\section{Introduction}

Tidal effects in binary systems of stars have long been investigated by many authors.
The primary star in a binary system is affected by the gravitational field of the companion star that orbits
around the primary, and vice versa.
The tides affect binary evolution, leading to synchronization between the orbital motion and 
stellar rotation, circularization of the binary orbit, and change of the orbital separation between the stars
(e.g., Hut, 1981).
It was a common practice to consider equilibrium and dynamical tides separately.
Equilibrium tides are tides considered in the limit of $\omega_{\rm tide}\rightarrow 0$,
where $\omega_{\rm tide}$ is the forcing frequency caused by the orbital motion of the companion.
Dynamical tides, on the other hand, are time dependent responses to the orbital motion of the companion.
In the case of dynamical tides, frequency resonance between $\omega_{\rm tide}$
and natural frequencies of the star can take place and is expected to have significant effects on the 
binary evolution.
It is dissipative processes accompanied by the tidal responses in binary stars that drive binary evolution.

Analytical and numerical studies of the tidal effects on binary evolution have been active 
since Zahn (1970, 1975, 1977) and Savonije \& Papaloizou (1983, 1984) tried
to estimate the time scales of synchronization and circularization of binary systems.
In these studies, the tidal potential due to the companion star was assumed to be a small perturbation
to the primary star so that the tidal responses of the primary are described by a linear theory of perturbations of stars.
The magnitudes of the tidal responses is proportional to the parameter $f_0\propto (M_2/a_{\rm orb}^3)(R^3/M)$,
where $M$ and $R$ are the mass and radius of the primary star, $M_2$ is the mass of the companion star, and
$a_{\rm orb}$ is the mean orbital separation between the stars.
The tidal responses also depend on the forcing frequency $\omega_{\rm tide}$ and attain very large amplitudes
when $\omega_{\rm tide}$ is in resonance with low frequency $g$-modes of the star.
In a linear theory of perturbations, the resonant amplitudes of tidal responses are limited by dissipations 
such as produced by non-adiabatic effects and/or viscous effects accompanied with the responses.
Although these early studies of tidal effects on binary systems of stars did not take account of the effects of stellar rotation on tidal responses, Savonije, Papaloizou, \& Alberts (1995), 
Savonije \& Papaloizou (1997), Witte \& Savonije (1999ab, 2001, 2002), Ogilvie \& Lin (2004, 2007)
numerically investigated tidal responses of rotating stars.
Stellar rotation brings about some complexities when estimating the tidal effects on binary stars and on the binary evolution.
Because of the Coriolis force as a restoring force
there appear rotational modes such as inertial modes and $r$-modes, whose frequencies are proportional to the rotation frequency $\Omega$ of the star (e.g., Unno et al 1989).
Inertial modes propagate in isentropic regions found in the convective regions of stars, while 
$r$-modes, which are retrograde modes, propagate in the radiative envelope.
If we consider tidal effects possibly caused by resonance between the forcing frequency $\omega_{\rm tide}$ and oscillation modes of rotating stars,
we have to take into consideration rotational modes as well as
$g$-modes when $\omega_{\rm tide}$ is comparable to or smaller than $\Omega$.
Witte \& Savonije (1999b, 2001), for example, discussed the effect of tidal locking on the binary evolution, and 
Ogilvie \& Lin (2004) numerically investigated tidal excitation of inertial modes of a giant planet that has a large convective core and a thin radiative envelope.

Oscillations of rotating stars may excite axisymmetric mean flows in the stars.
Lee et al. (2016) studied such mean flows driven by pulsationally unstable low frequency $g$- and $r$-modes
of slowly pulsating B (SPB) stars, using a theory of wave-mean flow interaction (see B\"uhler 2014 for
a review of the theory).
In SPB stars, numerous low frequency oscillation modes are excited by the 
opacity bump mechanism operating in the temperature regions of $T\sim 1.5\times10^5$K (e.g., Dziembowski et al 1993;
Gautschy \& Saio 1993).
Lee et al. (2016) have shown that self-excited low frequency oscillations drive axisymmetric mean flows and that
the $\phi$-component of the mean flow velocities dominates other components.
The velocities of the mean flows become large in the surface layers of the envelope
where non-adiabatic effects are significant.
Note that, for mean flows driven by pulsationally unstable low frequency modes, the amplitudes of the modes and mean flows are undetermined within a linear theory of oscillation, unless we take account of amplitude saturation mechanisms such as non-linear couplings between oscillation modes (e.g., Lee 2012).

In this paper, we investigate axisymmetric mean flows driven by tidal responses of the primary star in a binary system.
We treat the tidal responses, which are excited by orbital motion of the companion, 
as small amplitude perturbations of first order in the parameter $f_0$.
We assume that axisymmetric mean flows of the second order
are driven via non-linear effects of the responses.
To compute mean flows for uniformly rotating stars, we use the formulation given by Lee et al. (2016), who  
employed an Eulerian perturbation theory of second order, where zonal averaging
was used to pick up second order axisymmetric perturbations.
We calculate tidal responses and mean flows as a function of the forcing frequency $\omega_{\rm tide}$.

We use a zero-age-main-sequence (ZAMS) star model of $15M_\odot$ as the background model
for mean flow calculations.
The ZAMS model has a chemically homogeneous and rather simple structure composed of a convective core and a radiative envelope and
has a simple frequency spectrum of low frequency oscillation modes, which are all expected to be pulsationally stable.
We also assume uniform and slow rotation of the star just for simplicity. 
For rapidly rotating stars, the frequency ranges of low radial order $g$-modes and inertial modes overlap,
which would make the analyses more complicated.
In \S 2, we give a brief description of the formulation we use for tidal response calculations 
and show some numerical results of the responses for the $15M_\odot$ ZAMS model.
In \S 3, we describe numerical results for tidally driven axisymmetric mean flows.
We conclude in \S 4.

\section{Calculation of tidal responses}

\subsection{Basic equations for tidally perturbed stars}

In a binary system of stars, the orbital motion of the companion star excites via tidal potential $\Psi$ time dependent tidal responses
in the primary star, and vice versa.
We let $\omega_{\rm tide}$ denote the forcing frequency associated with the tidal potential $\Psi$.
If the tidal potential $\Psi$ is treated as a small perturbation to the primary star,
the governing equations for tidal responses in the primary are given by a set of linearized
basic equations of fluid dynamics:
\be
{\partial\pmb{v}'\over\partial t}+2\pmb{\Omega}\times\pmb{v}'=-{1\over\rho}\nabla p'+{\rho'\over\rho^2}\nabla p-\nabla\left(\Phi'+\Psi\right),
\label{eq:linearizedeom}
\ee
\be
{\partial\rho'\over\partial t}+\nabla\cdot\left(\rho\pmb{v}'\right)=0,
\ee
\be
\nabla^2\Phi^\prime=4\pi G\rho^\prime,
\ee
\be
\rho T\left({\partial s'\over\partial t}+\pmb{v}'\cdot\nabla s\right)=\left(\rho\epsilon\right)'-\nabla\cdot\pmb{F}',
\ee
\be
\pmb{F}^\prime_{\rm rad}=-\lambda_{\rm rad}^\prime\nabla T-\lambda_{\rm rad}\nabla T',
\label{eq:eqradtrans}
\ee
where
$
\lambda_{\rm rad}={(4ac/ 3)}{(T^3/\kappa\rho)}
$
is the radiative conduction coefficient,
$\pmb{v}$ is the fluid velocity,
$p$ is the pressure,
$\rho$ is the mass density, $T$ is the temperature,
$s$ is the specific entropy, $\pmb{F}$ is the energy flux, 
$\Phi$ is the gravitational
potential, $\Psi$ is the tidal potential,
$\epsilon$ is the  nuclear energy generation rate per gram,  $\kappa$ is the opacity, 
$G$ is the gravitational constant, $a$ is the radiation constant, $c$ is the velocity of light, and
the primed quantities indicate Eulerian perturbations.
Here, the companion star is assumed to be in the equatorial plane of the primary star, and 
the angular velocity of rotation $\pmb{\Omega}$ of the primary is assumed constant and parallel to
the normal to the orbital plane.
We have also assumed that the hydrostatic equilibrium in the primary star is given by
$
\nabla p=-\rho\nabla\Phi,
$
that is, we have ignored equilibrium deformations due to rotation and tides.
The energy flux $\pmb{F}$ is given by $\pmb{F}=\pmb{F}_{\rm rad}$ in the radiative regions
and $\pmb{F}=\pmb{F}_{\rm rad}+\pmb{F}_{\rm conv}$ in the convective regions, where
$\pmb{F}_{\rm rad}$ and $\pmb{F}_{\rm conv}$ are the radiative and convective energy fluxes, respectively. 
For perturbations of the convective energy flux $\pmb{F}_{\rm conv}$, we assume $\delta\left(\nabla\cdot\pmb{F}_{\rm conv}\right)=0$
(see, e.g., Unno et al 1989),
where $\delta$ indicates the Lagrangian perturbation.

In this paper, we assume that the time dependence of the perturbations is given by the factor $e^{\rmi\omega t}$ with $\omega$ being the oscillation frequency observed in the co-rotating frame of the star.
For uniformly rotating stars, the Euler perturbations of the velocity, $\pmb{v}'$, is given by
\be
\pmb{v}'=\rmi\omega\pmb{\xi},
\ee
where $\pmb{\xi}=\xi_r\pmb{e}_r+\xi_\theta\pmb{e}_\theta+\xi_\phi\pmb{e}_\phi$ is the displacement vector given in spherical polar coordinates $(r,\theta,\phi)$, and
$\pmb{e}_r$, $\pmb{e}_\theta$, and $\pmb{e}_\phi$ are the orthonormal vectors in the $r$, $\theta$, and $\phi$
directions, respectively.

\subsection{calculating equilibrium tide}


As a response to the tidal potential $\Psi$, equilibrium tides may be defined as
(e.g., Ogilvie \& Lin 2004; see also Goldreich \& Nicholson 1989)
\be
\xi_{r,\rm e}=-{\Phi'_{\rm e}+\Psi\over g}, \quad \xi_{h,\rm e}={1\over l(l+1)r}{d\over dr}r^2\xi_{r,\rm e},
\label{eq:xire}
\ee
\be
\rho_{\rm e}'=-\xi_{r,\rm e}{d\rho\over dr}, \quad p'_{\rm e}=-\xi_{r,\rm e}{dp\over dr},
\label{eq:rhoprimee}
\ee
and
\be
\nabla^2\Phi_{\rm e}'=4\pi G\rho_{\rm e}', 
\label{eq:poissone}
\ee
where $g=d\Phi/dr$.
Note that, if we write $\pmb{\xi}_{\rm e}=\left(\xi_{r,\rm e}\pmb{e}_r+\xi_{h,\rm e}\nabla\right)Y_l^me^{\rmi\omega_{\rm tide}t}$,
we have
\be
\nabla\cdot\pmb{\xi}_{\rm e}=0,
\ee
indicating that the equilibrium tide is incompressible.
Making use of equations (\ref{eq:xire}), (\ref{eq:rhoprimee}), and (\ref{eq:poissone}), we obtain
\be
\nabla^2\left(\Phi_{\rm e}'+\Psi\right)={4\pi G\over g}{d\rho\over dr}\left(\Phi_{\rm e}'+\Psi\right),
\label{eq:poissonpsi}
\ee
where we have assumed $\nabla^2\Psi=0$.
Integrating the differential equation ({\ref{eq:poissonpsi}) with appropriate boundary conditions, we obtain equilibrium tides $\Phi_{\rm e}'$,
$\pmb{\xi}_{\rm e}$, $\rho_{\rm e}'$ and $p_{\rm e}'$.



We are interested in tidal responses excited by the potential given, in an inertial frame, as
\be
\Psi
= -f_0{GM\over R}x^2Y_2^{-2}e^{2\rmi\Omega_{\rm orb}t}\equiv \Psi_2Y_2^{-2}e^{2\rmi\Omega_{\rm orb}t},
\label{eq:tidalpot}
\ee
where $M$ and $R$ are the mass and radius of the primary star, $x=r/R$, $\Omega_{\rm orb}=\sqrt{G(M+M_2)/a_{\rm orb}^3}$ is the mean angular velocity of the orbital motion, $a_{\rm orb}$ is the mean separation between the primary and companion
stars, and
\be
f_0=\sqrt{6\pi\over 5}{GM_2/a_{\rm orb}^3\over\sigma_0^2}=\sqrt{6\pi\over 5}{q\over \left({a_{\rm orb}/ R}\right)^{3}},
\ee
where $M_2$ is the mass of the companion star, $q=M_2/M$, and 
$
\sigma_0=\sqrt{GM/R^3}.
$
For $l=2$, assuming $\Phi'_e \propto Y_2^{-2}e^{2\rmi\Omega_{\rm orb}t}$,
equation (\ref{eq:poissonpsi}) may reduce to (see, e.g., Schwarzschild 1958)
\be
{1\over x^2}{d\over dx}\left(x^2{d\over dx}F\right)
-\left({6\over x^2}-k^2\right)F=0,
\label{eq:difeqf}
\ee
where
\be
F={\Phi_{\rm e}'+\Psi\over GM/R}, \quad k^2=-R^2{4\pi G\over g}{d\rho\over dr}.
\ee

We integrate the second order ordinary differential equation (\ref{eq:difeqf}) from the centre to the surface of the star applying the boundary conditions given below.
The boundary condition at the centre is the regularity condition given by 
\be
{F}-{x_0^2-k_0^2x_0^4/14\over 2x_0-(2/7)k_0^2x_0^3}{dF\over dx}=0,
\ee
where $x_0\ll1$ and $k_0$ is the value of $k^2$ at $x=x_0$.
The surface boundary condition at $x=1$ is given by (e.g., Ogilvie \& Lin 2004)
\be
{d\ln\left|\Phi_e'\right|\over d\ln r}=-3,
\ee
which leads to
\be
3F+{dF\over dx}=-5f_0.
\ee
The amplitudes of the equilibrium and dynamical tides are proportional to the
parameter $f_0$.

\subsection{equations for tidal responses}

Under the Cowling approximation, neglecting the Eulerian perturbation of the gravitational potential, 
we write the linearized equation of motion (\ref{eq:linearizedeom}) as
\be
-\rho\omega^2\pmb{\xi}+2\rmi\omega\Omega\rho\pmb{\Omega}\times\pmb{\xi}=-\nabla p'-{\rho'}\nabla \Phi-\rho\nabla\left(\Phi'_e+\Psi\right).
\label{eq:linearizedeom2}
\ee
The perturbed continuity and entropy equations are
\be
\rho'+\nabla\cdot\left(\rho\pmb{\xi}\right)=0,
\label{eq:eqcontinuity}
\ee
\be
\rmi \omega\rho T\delta s=\left(\rho\epsilon\right)'-\nabla\cdot\pmb{F}',
\label{eq:eqentropy}
\ee
and the perturbed equation of state is given by
\be
{\rho'\over\rho}=-rA{\xi_r\over r}+{1\over\Gamma_1}{ p'\over p}-\alpha_T{\delta s\over c_p},
\label{eq:eos}
\ee
where
\be
rA={d\ln\rho\over d\ln r}-{1\over\Gamma_1}{d\ln p\over d\ln r},
\ee
and
\be
\Gamma_1=\left({\partial\ln p\over\partial\ln\rho}\right)_{\rm ad}, \quad \alpha_T=-\left({\partial \ln\rho\over\partial\ln T}\right)_p.
\ee

Since separation of variables is not possible for the perturbations in rotating stars,
we use finite series expansions in terms of spherical harmonic functions $Y_l^m(\theta,\phi)$ to 
represent the perturbations.
Assuming that the equilibrium state is axisymmetric about the rotation axis, 
we expand the three components of the displacement vector $\pmb{\xi}(\pmb{x},t)$ as
\be
{\xi_r}=r\sum_{j=1}^{j_{\rm max}}S_{l_j}(r)Y_{l_j}^m(\theta,\phi)e^{\rmi\omega t},
\label{eq:xiexp_r}
\ee
\be
{\xi_\theta}=r\sum_{j=1}^{j_{\rm max}}\left[H_{l_j}(r)\frac{\partial}{\partial\theta}
Y_{l_j}^m(\theta,\phi)+T_{l'_j}(r)\frac{1}{\sin\theta}\frac{\partial}{\partial\phi}Y_{l'_j}^m(\theta,\phi)
\right]e^{\rmi\omega t},
\label{eq:xiexp_theta}
\ee
\be
{\xi_\phi}=r\sum_{j=1}^{j_{\rm max}}\left[H_{l_j}(r)\frac{1}{\sin\theta}\frac{\partial}{\partial\phi}
Y_{l_j}^m(\theta,\phi) - T_{l'_j}(r)\frac{\partial}{\partial\theta}Y_{l'_j}^m(\theta,\phi)
\right]e^{\rmi\omega t},
\label{eq:xiexp_phi}
\ee
and the Eulerian pressure perturbation,
$p^\prime(\pmb{x},t)$, as
\be
p^\prime=\sum_{j=1}^{j_{\rm max}}p^\prime_{l_j} (r)Y_{l_j}^m\left(\theta,\phi\right)e^{\rmi\omega t},
\label{eq:pexpansion}
\ee
where $l_j=2(j-1)+|m|$ and $l'_j=l_j+1$ for even modes, and $l_j=2j-1+|m|$ and
$l'_j=l_j-1$ for odd modes for $j=1~,2~,3\cdots,~j_{\rm max}$ (e.g., Lee \& Saio 1987).
As indicated by the expressions given above, the perturbations are proportional to the factor $e^{\rmi m\phi+\rmi\omega t}$,
and if we let $\sigma$ denote the oscillation frequency observed in an inertial frame,
the oscillation frequencies $\omega$ in the co-rotating frame is given by $\omega=\sigma+m\Omega$.
Substituting the series expansions of the perturbations into the perturbed basic equations (\ref{eq:linearizedeom2}) to (\ref{eq:eos}) and (\ref{eq:eqradtrans}), we obtain
a finite set of linear ordinary differential equations for the expansion coefficients (see the Appendix).
For a given tidal potential $\Psi$ and for a tidal forcing frequency $\omega=\omega_{\rm tide}$, we solve the finite set of differential equations  
with boundary conditions imposed at the centre and the surface of the star.
The inner boundary conditions are the regularity condition for the perturbations and 
the condition for adiabatic oscillation given by $\delta s=0$.
The outer boundary conditions are given by $\delta p=0$ and $\delta L_{\rm rad}=\delta\left(4\pi R^2\sigma_{\rm SB}T^4\right)$ with $\sigma_{\rm SB}$ being the Stefan-Boltzmann constant.
See the Appendix for the detail.

\begin{figure}
\begin{center}
\resizebox{0.45\columnwidth}{!}{
\includegraphics{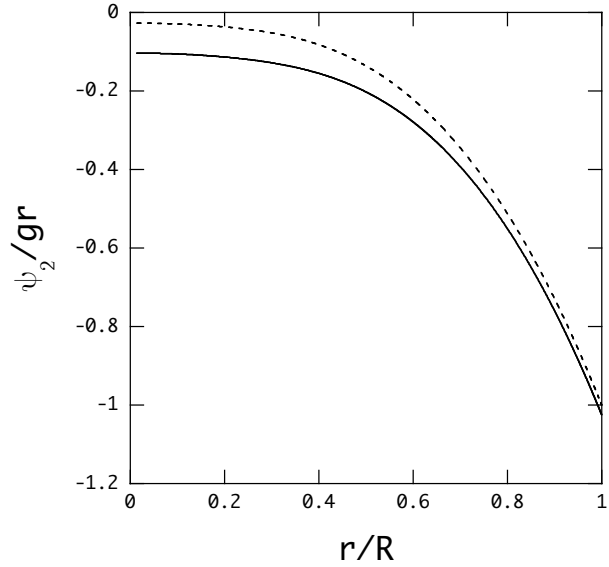}}
\end{center}
\caption{Tidal potentials $\psi_{2}/gr=(\Phi'_{e,2}+\Psi_2)/gr$ (solid curve) and $\Psi_2/gr$ (dotted curve) for the $15M_\odot$ ZAMS model where $g=GM_r/r^2$ and $M_r=\int_0^r4\pi r^2\rho dr$.}
\end{figure}

\subsection{calculating tidal responses}

To investigate tidal responses of massive stars, we use 
a $15M_\odot$ zero age main sequence (ZAMS) model
computed with a standard stellar evolution code
using the OPAL opacity (Iglesias \& Rogers 1996) for $X=0.7$ and $Z=0.02$.
For this model, we plot 
$\psi_{2}/gr=(\Phi'_{e,2}+\Psi_2)/gr$ (solid line) and $\Psi_2/gr$ (dotted line) for $f_0=1$ in Figure 1.
The difference between $\psi_{2}$ and $\Psi_2$ gives the contribution of the equilibrium tides $\Phi^\prime_e$,
which become significant in the stellar core.
Note that $(\xi_r/r)_e$ of the equilibrium tide is given by $(\xi_r/r)_e=-\psi_{2}/gr$.

We define the tidal torque ${\cal T}_{l}$ for $l=2$ as (e.g., Savonije \& Papaloizou 1984)
\be
{\cal T}_{2}=-\int dV\overline{\rho'_2{\partial\Phi'_2\over\partial\phi}}={m\over 2}\int_0^R \rho gr^3dr {\rm Im}\left({\rho'^*_2\over\rho}{\Phi'_{e,2}+\Psi_2\over gr}\right)\equiv {GM^2\over R}\overline{\cal T}_2,
\label{eq:tidaltorque}
\ee
where
\be
\overline{f}={1\over 2\pi}\int_0^{2\pi}fd\phi,
\label{eq:zonalav}
\ee
and for a product of the first order perturbations $f_1$ and $f_2$, which are complex quantities, we may evaluate
\be
\overline{f_1f_2}={1\over 2}{\rm Re}\left(f_1^*f_2\right)={1\over 2}{\rm Re}\left(f_1f_2^*\right),
\ee
where the asterisk $(^*)$ indicates complex conjugation.

In Figure 2 we plot the normalized torque $\left|\overline{{\cal T}_2}\right|$ as a function of the forcing frequency $\bar\omega_{\rm tide}$ for the $15M_\odot$ model for $\bar\Omega=0.1$ (left panel) and for $\bar\Omega=0.4$ (right panel), where $\bar\omega_{\rm tide}$ and $\bar\Omega$ denote dimensionless frequencies defined as
$\bar\omega_{\rm tide}=\omega_{\rm tide}/\sigma_0$ and $\bar\Omega=\Omega/\sigma_0$.
Since perturbations are assumed to be proportional to $e^{\rmi(\omega t+m\phi)}$ in this paper,
the positive (negative) frequency $\omega_{\rm tide}$ corresponds to prograde (retrograde) forcing observed
in the co-rotating frame of the star.
As shown by the figure, there appear numerous peaks, produced by resonance between the forcing frequency $\omega_{\rm tide}$
and natural frequencies of $g$-modes and inertial modes of the star.
The ZAMS model have a convective core and a radiative envelope, and $g$-modes propagate in the radiative envelope
and inertial modes in the convective core where we have $N^2\approx 0$ with $N$ being the Brunt-V\"ais\"al\"a frequency.
On the negative side of $\bar\omega_{\rm tide}$, we also find a sequence of resonance peaks associated with $r$-modes, which are retrograde modes propagating in the 
radiative envelope and have frequencies
$\bar\omega\gtsim 2m\bar\Omega/l'(l'+1)\approx -0.033$ for $\bar\Omega=0.1$ and $\bar\omega\gtsim-0.1333$ for $\bar\Omega=0.4$ when $m=-2$ and $l'=3$.
As shown by the figure, the tidal torque is significantly reduced in the inertial regime of $\left|\omega/\Omega\right|\le2$,
except for that caused by the $r$-modes.
For rapidly rotating stars, this inertial frequency regime overlaps the frequency ranges of low radial order $g$-modes.
Although most of the conspicuous peaks result from resonance with $l=-m=2$ $g$-modes,
we also find sequences of less pronounced peaks, which are produced by resonance with $g$-modes of $l=4$ and $m=-2$.
For the $15M_\odot$ model, the tidal torques ${\cal T}_{2}$ has opposite signs between prograde and retrograde forcing
and the sign stays the same as a function of $\omega_{\rm tide}$ except for very low frequency regions.
For comparison, we tabulate the complex eigenfrequency $(\bar\omega_{\rm R},\bar\omega_{\rm I})$ of low radial order $l=-m=2$ $g$-modes of the $15M_\odot$ model for $\bar\Omega=0.1$, where $\bar\omega_{\rm I}>0$ indicates that the mode is pulsationally stable.

\begin{table*}
\begin{center}
\caption{Complex eigenfrequency $\bar\omega=(\bar\omega_{\rm R},\bar\omega_{\rm I})$ of low radial order $l=-m=2$ $g$-modes of 
the $15M_\odot$ ZAMS model for $\bar\Omega=0.1$.}
\label{symbols}
\begin{tabular}{@{}crrrr}
\hline
    &  prograde & & retrograde & \\
\hline
mode & $ \ \bar\omega_{\rm R}$ & $ \ \bar\omega_{\rm I}$ & $ \ \bar\omega_{\rm R}$ & $ \ \bar\omega_{\rm I}$ \\
\hline
$g_1\cdots\cdots$ & 1.42585 & 2.39E-8 & -1.45273 & 2.38E-8 \\
$g_2\cdots\cdots$ & 0.91627 & 6.95E-8 & -0.95505 & 8.24E-8 \\
$g_3\cdots\cdots$ & 0.67962 & 2.36E-7 & -0.72648 & 2.85E-7 \\
$g_4\cdots\cdots$ & 0.53311 & 1.03E-6 & -0.58549 & 1.00E-7 \\
$g_5\cdots\cdots$ & 0.43741 & 5.66E-6 & -0.49325 & 3.96E-6 \\
$g_6\cdots\cdots$ & 0.37045 & 2.06E-5 & -0.42877 & 1.63E-5 \\
$g_7\cdots\cdots$ & 0.32112 & 8.12E-5 & -0.38129 & 5.81E-5 \\
$g_8\cdots\cdots$ & 0.28513 & 2.97E-4 & -0.34638 & 2.21E-4 \\
$g_9\cdots\cdots$ & 0.25756 & 6.64E-4 & -0.32027 & 5.39E-4 \\
$g_{10}\cdots\cdots$ & 0.23380 & 1.04E-3 & -0.29800 & 3.56E-4 \\
\hline
\end{tabular}
\medskip
\end{center}
\end{table*}


\begin{figure}
\begin{center}
\resizebox{0.45\columnwidth}{!}{
\includegraphics{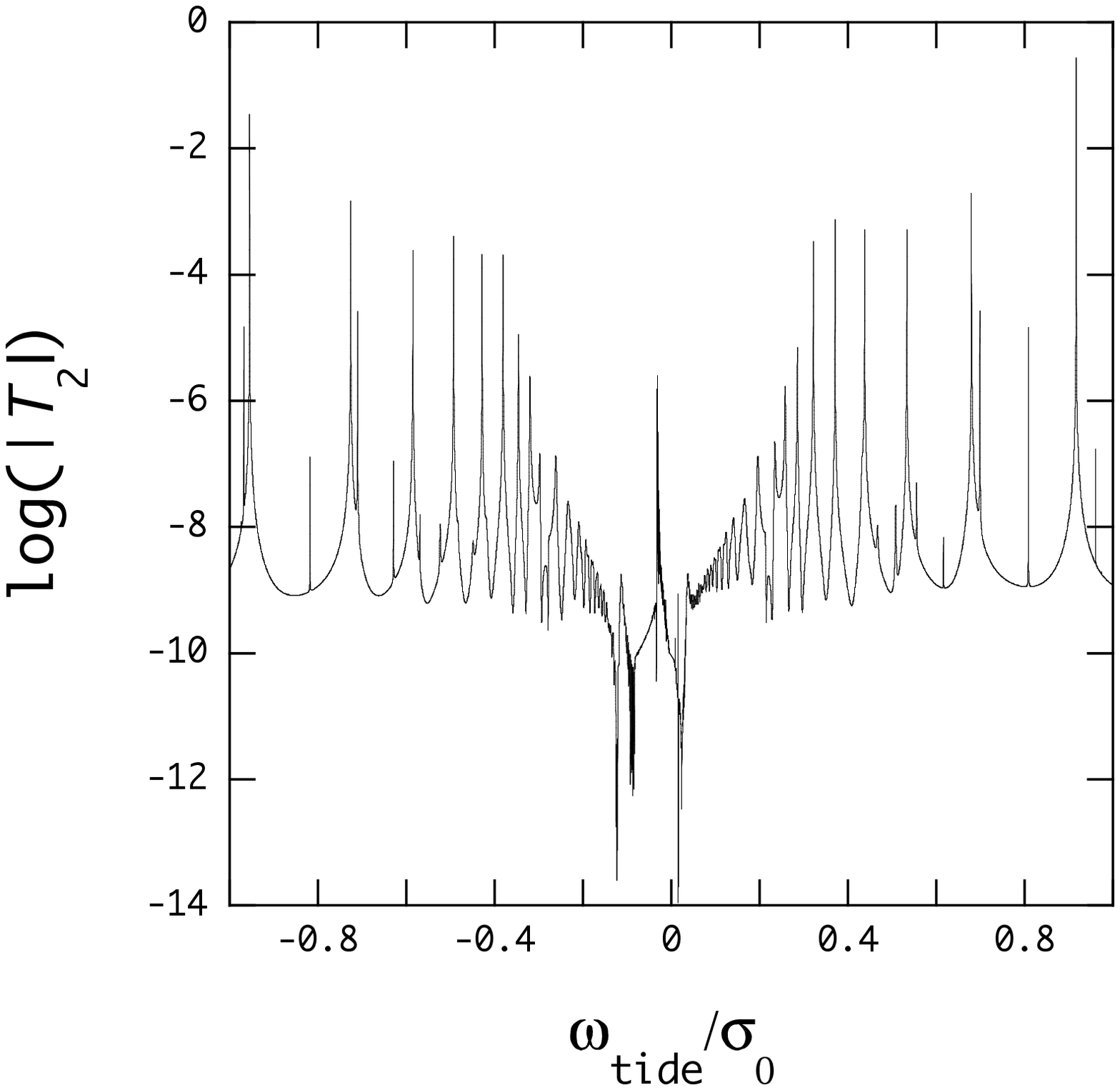}}
\resizebox{0.45\columnwidth}{!}{
\includegraphics{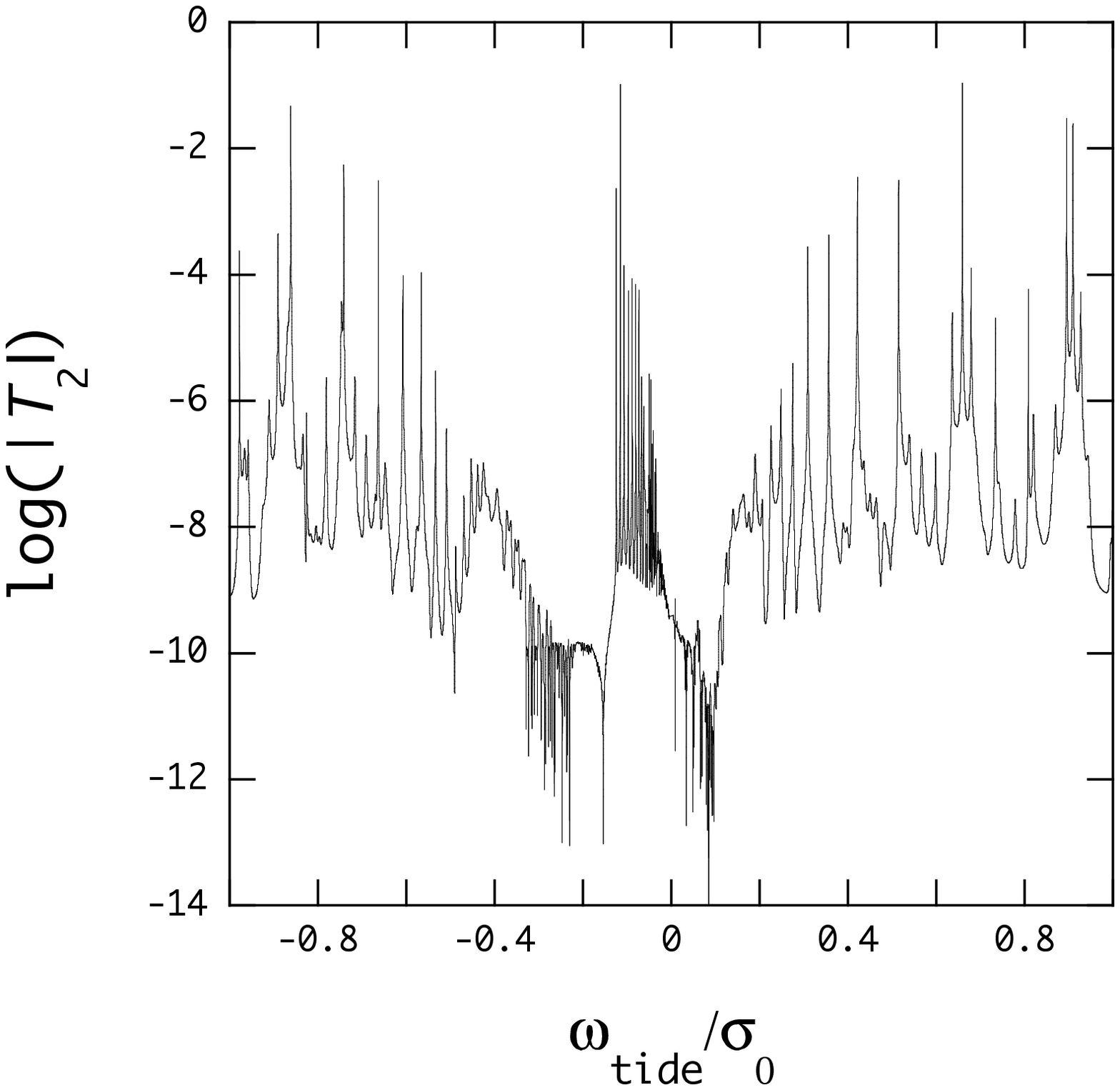}}
\end{center}
\caption{Tidal torque $\left|\overline{\cal T}_2\right|$ versus the forcing frequency $\bar\omega_{\rm tide}=\omega_{\rm tide}/\sigma_0$ for $\bar\Omega=0.1$ (left panel)
and $\bar\Omega=0.4$ (right panel) for the
$15M_\odot$ ZAMS model, where we use $j_{\rm max}=15$ and $f_0=1$.
}
\label{fig:t2}
\end{figure}

Figure 3 shows tidal responses of the $15M_\odot$ model at $\bar\omega_{\rm tide}=0.3704823794$ (left panel) and at $\bar\omega_{\rm tide}=0.35$ (right panel), where the real part of the expansion coefficient $S_{l}$ 
is plotted for $l=2$, 4, and 6.
The left panel gives an example of tidal responses at a forcing frequency $\omega_{\rm tide}$ in resonance with a $l=-m=2$ $g$-mode, while the right panel
shows a tidal response in off-resonance with low frequency modes, in which the response $S_{l=2}$ is approximately given by the equilibrium tide $(\xi_r/r)_e=-\psi_2/gr$ as shown by the long dashed line.
In both cases, the component $S_{l=2}$ is dominating because the tidal potential $\Psi$ is here proportional to $Y_{l=2}^{m=-2}$.
The amplitude at resonance can be much larger than the amplitude in off-resonance, which
is comparable to $(\xi_r/r)_e$.

\begin{figure}
\begin{center}
\resizebox{0.45\columnwidth}{!}{
\includegraphics{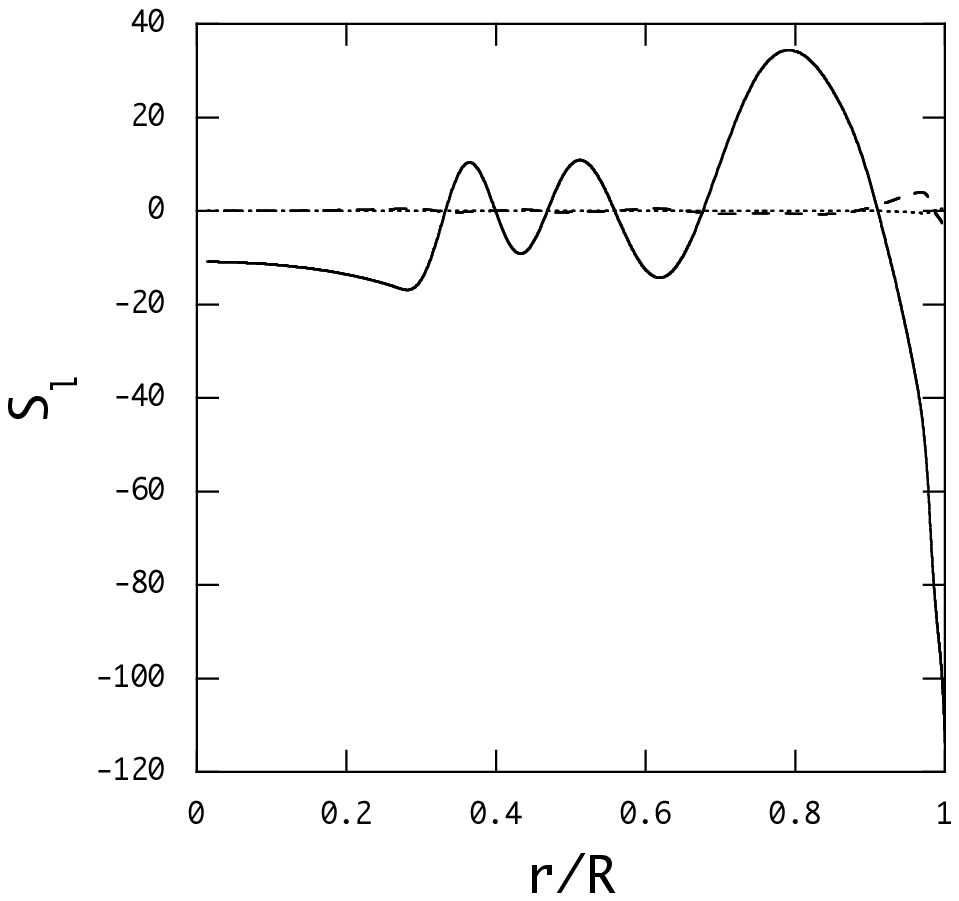}}
\resizebox{0.45\columnwidth}{!}{
\includegraphics{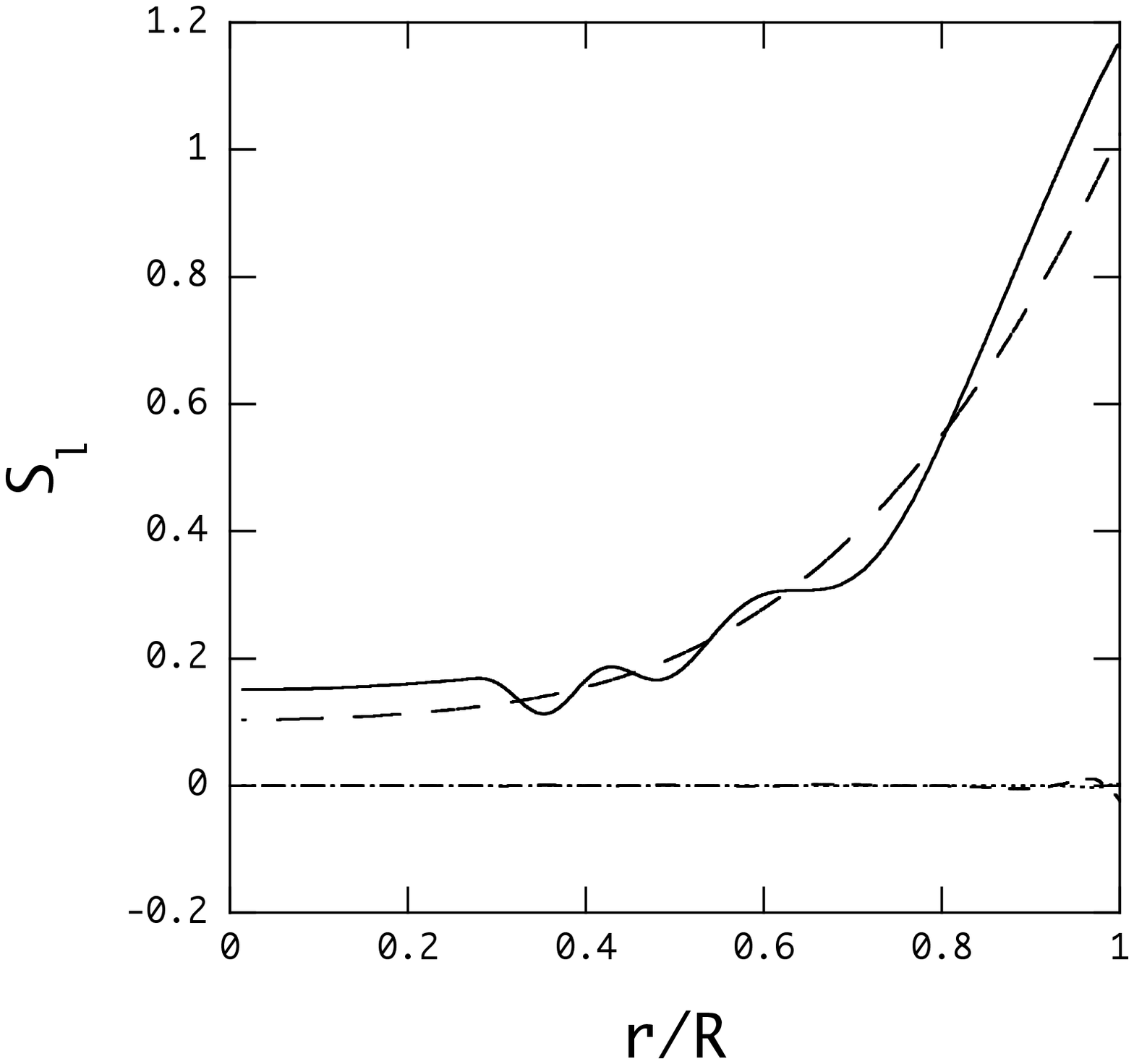}}
\end{center}
\caption{Tidal responses $S_{l}$ at $\bar\omega_{\rm tide}=0.3704823794$ (left panel)
and at $\bar\omega_{\rm tide}=0.345$ (right panel), where the solid, dashed and dotted curves are for 
the expansion coefficients $S_2$, $S_4$, and $S_6$, respectively.
The left panel shows the responses in
resonance with the $l=|m|=2$ $g_6$ mode and the right panel shows the responses in off-resonance with low frequency modes.
The long dashed line in the right panel depicts $(\xi_{r}/r)_e=-\psi_2/gr$, that is,
the equilibrium tide for the forcing frequency.
Here, we use $15M_\odot$ ZAMS model and assume $\bar\Omega=0.1$ and $f_0=1$.
}
\label{fig:tidalresponses}
\end{figure}

For comparison, we plot in Figure \ref{fig:freegmode} the eigenfunction $S_l$ of the $l=-m=2$ $g_6$-mode
and the derivative $dw/dr$ of the work function $w$
defined as (e.g., Unno et al 1989)
\be
w(r)=-\pi \int_0^r \alpha_T{\rm Im}\left(\delta p^{*}{\delta s\over c_p}\right) r^2dr,
\label{eq:workfunction}
\ee
where $\bar\Omega=0.1$ is assumed.
We normalize the eigenfunction by $S_{l_1}=1$ at the surface
Note that $dw/dr>0$ ($dw/dr<0$) indicates excitation (damping) regions for an oscillation mode.
The $g_6$-mode is pulsationally stable, 
that is, the amount of damping exceeds that of driving in the interior.
Comparing the left panels of Figures 3 and 4, we find that the tidal response $S_l$ at the resonance looks quite similar to the eigenfunction $S_l$, except for the amplitudes.
The plot of $dw/dr$ in Figure 4 indicates that there extend an excitation region for the mode below $x=r/R\sim 0.94$, above which non-adiabatic damping prevails up to the surface.
 
\begin{figure}
\begin{center}
\resizebox{0.45\columnwidth}{!}{
\includegraphics{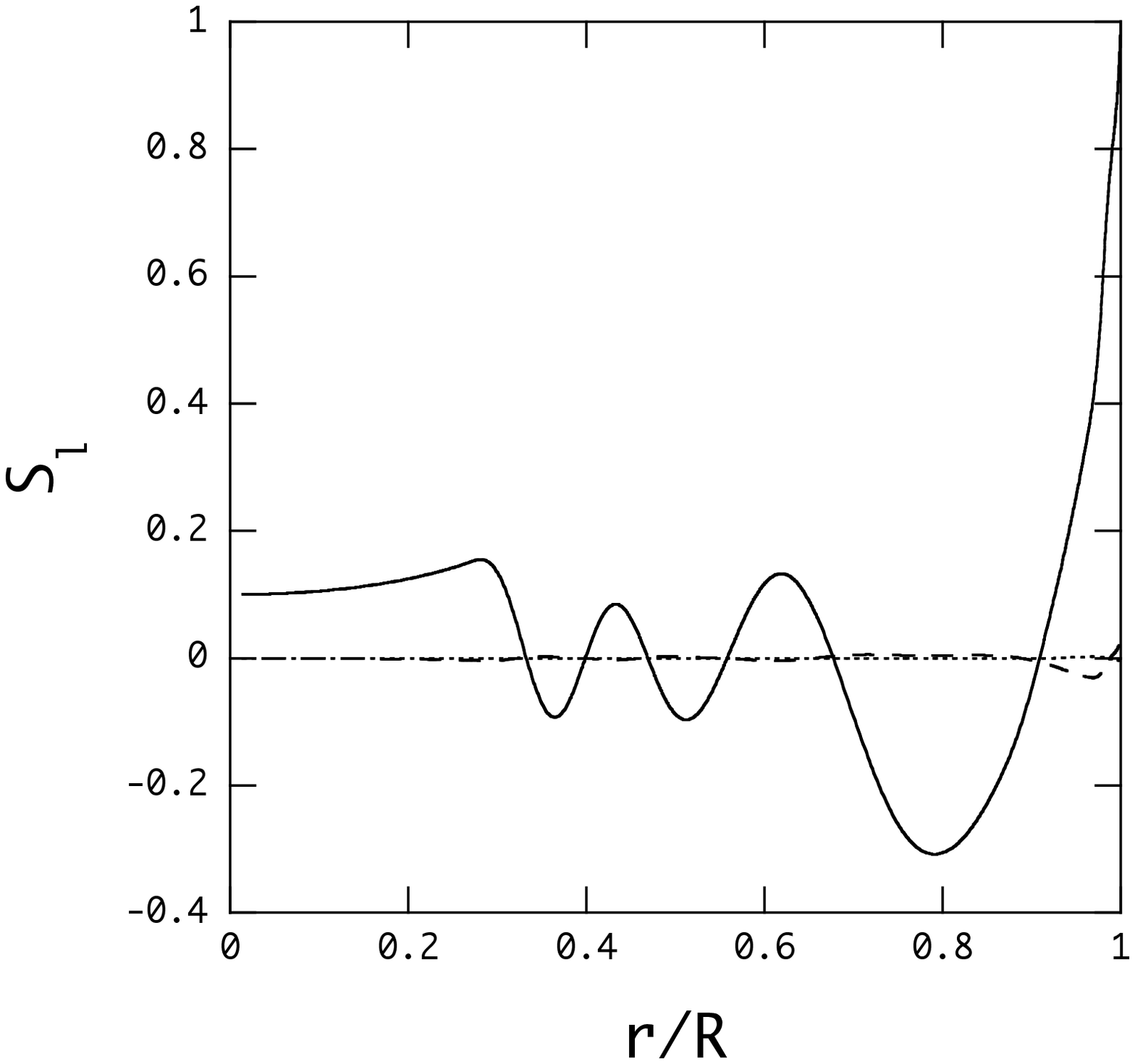}}
\resizebox{0.45\columnwidth}{!}{
\includegraphics{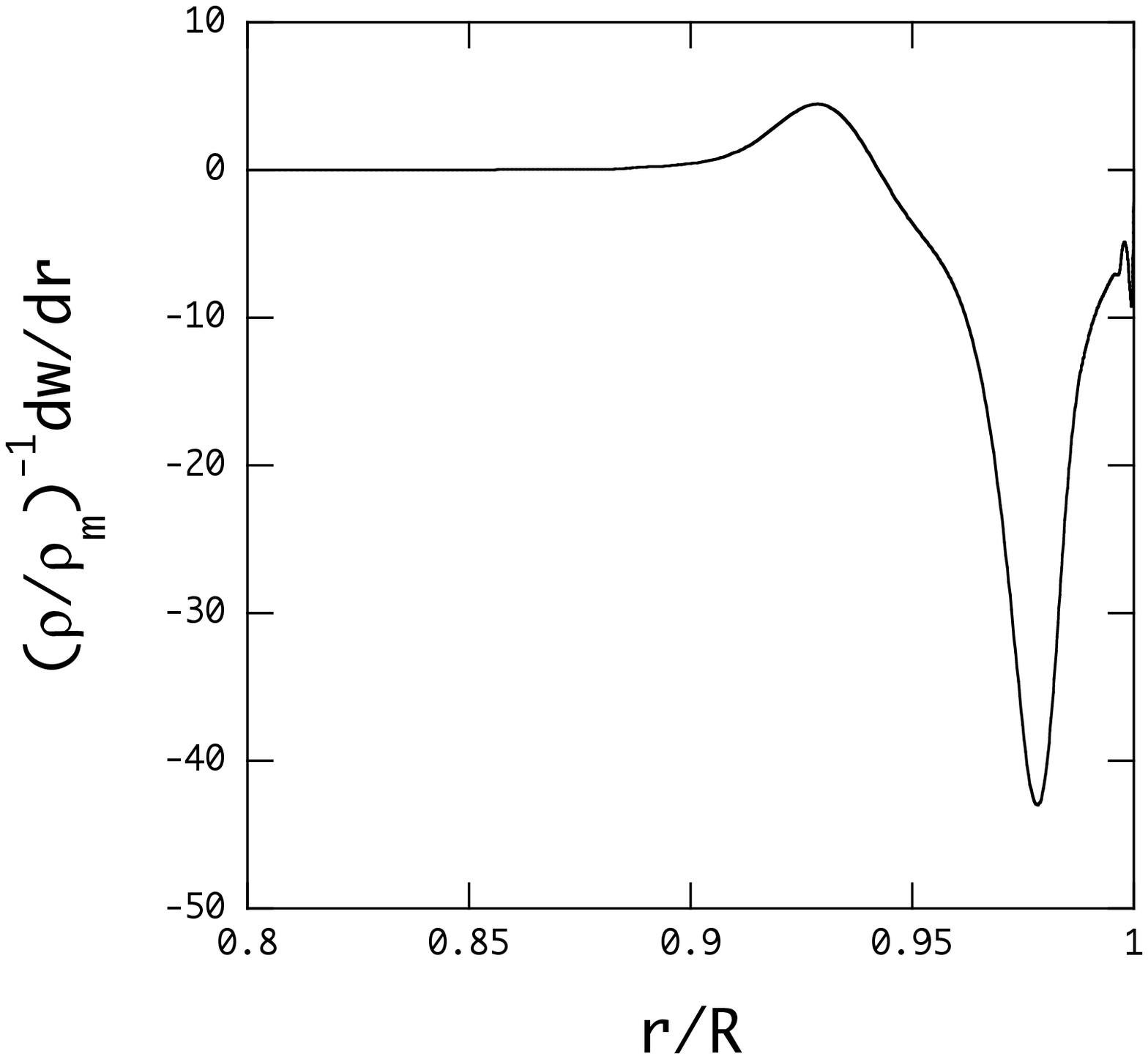}}
\end{center}
\caption{Eigenfunction $S_{l}$ of the $l=-m=2$ $g_6$-mode of $\bar\omega=0.3704512$ (left panel)
and the derivative $(\rho/\rho_m)^{-1}dw/dr$ of the work function $w$ (right panel) as a function of $x=r/R$, where we use the $15M_\odot$ ZAMS model at $\bar\Omega=0.1$, and $\rho_m=M/(4\pi R^3/3)$ is the mean density of the star.
Amplitude normalization is given by $S_{2}=1$ at the surface.
In the left panel, the solid, dashed and dotted curves
are for the expansion coefficients $S_{2}$, $S_4$, and $S_6$, respectively.
}
\label{fig:freegmode}
\end{figure}

\section{Tidally Driven Mean Flows}

We calculate mean flows driven by tidal responses, using the formulation given by Lee et al (2016).
Here, tidal responses are considered as first order perturbations,
while mean flows as second order perturbations in the parameter $f_0$.

\subsection{perturbed equations of second order for mean flows}

When tidal responses have
small amplitudes and are regarded as a perturbation, any physical quantities $f\left(\pmb{x},t\right)$ of the primary star may be
represented by
\be
f(\pmb{x},t)=f^{(0)}(\pmb{x})+f^{(1)}(\pmb{x},t)+f^{(2)}(\pmb{x},t) +\cdots,
\ee
where $f^{(0)}$ denotes the equilibrium quantities, $f^{(1)}$ the Euler
perturbations of first-order, and $f^{(2)}$ the Eulerian perturbations of
second-order in $f_0$. 
Similarly, the velocity field
$\pmb{v}\left(\pmb{x},t\right)$ may be expanded as
\be
\pmb{v}(\pmb{x},t)=\pmb{v}^{(0)}(\pmb{x})+\pmb{v}^{(1)}(\pmb{x},t)+\pmb{v}^{(2)}(\pmb{x},t)+\cdots,
\ee
and the equilibrium state is assumed to be that of a uniformly rotating star so
that, in spherical polar coordinates $(r,\theta,\phi)$,
\be
\pmb{v}^{(0)}=r\sin\theta\Omega\pmb{e}_\phi,
\ee
where $\Omega$ is the angular velocity of rotation and assumed to be constant,
and $\pmb{e}_\phi$ is the unit vector in the azimuthal direction. 
In this paper, we ignore equilibrium deformation caused by rotation and tidal force and assume that $f^{(0)}$
depends only on the radial distance $r$ from the center of the star. 
We apply the Cowling approximation to second order perturbations, neglecting the Euler perturbation
$\Phi^{(2)}$ of the gravitational potential $\Phi$.

We employ a theory of wave-mean flow interaction to discuss axisymmetric flows
driven by tidal responses in rotating stars (e.g., B\"uhler 2014). We regard the
axisymmetric flows as mean flows, which contain both zero-th order and second order perturbations
in $f_0$.
The zero-th order quantities $f^{(0)}$ are those of equilibrium state, which is independent of time $t$.
The first order quantities $f^{(1)}$ are tidal responses of the primary star.
The second order quantities $f^{(2)}$ carry the time dependence of the mean flow.
To derive governing equations for the second-order perturbations $f^{(2)}$ for mean flows, we use the zonal
averaging defined by equation (\ref{eq:zonalav})
and, assuming
$
\overline {f^{(1)}}=0,
$
we obtain
\be
\overline{f}=\overline{f^{(0)}}+\overline{f^{(2)}}.
\ee
Here, we have ignored higher order terms $f^{(k)}$ with $k\ge 3$. Hereafter, we
simply write $f^{(0)}$ and $f^{(2)}$ respectively for $\overline{f^{(0)}}$ and
$\overline{f^{(2)}}$. The zonal averaging makes $f^{(0)}$ and $f^{(2)}$
independent of $\phi$.

We assume that non-oscillatory fluid flows arise in rotating stars
via nonlinear effects of tidal responses $f^{(1)}$.
Applying the zonal averaging to the basic equations,
we obtain a set of differential equations that govern
the second-order perturbations (Lee et al 2016): 
\be
\frac{{\partial \pmb{v} ^{\left( 2 \right)} }}{{\partial t}} + \pmb{v} ^{\left( 0 \right)}  \cdot \nabla \pmb{v} ^{\left( 2 \right)}  + \pmb{v} ^{\left( 2 \right)}  \cdot \nabla \pmb{v} ^{\left( 0 \right)}  + \frac{1}{{\rho ^{\left( 0 \right)} }}\nabla p^{\left( 2 \right)}  
+ g\frac{{\rho ^{\left( 2 \right)} }}{{\rho ^{\left( 0 \right)} }}
\pmb{e}_r  
=  - \overline {\pmb{v} ' \cdot \nabla \pmb{v} '}  + g\overline {\left( {\frac{{\rho '}}{{\rho ^{\left( 0 \right)} }}} \right)^2}  
\pmb{e}_r 
+ \frac{1}{{\rho ^{\left( 0 \right)} }}\overline {\frac{{\rho '}}{{\rho ^{\left( 0 \right)} }}\nabla p'},
\label{eq:secondmomcon}
\ee
\be
\frac{\partial\rho^{(2)}}{\partial t}+\nabla\cdot\left(\rho^{(0)}\pmb{v}^{(2)}\right)=-\overline{\nabla\cdot\left(\rho^\prime
\pmb{v}^\prime\right)},
\label{eq:secondcont}
\ee
\begin{eqnarray}
\frac{\partial s^{(2)}}{\partial t}+\pmb{v}^{(2)}\cdot\nabla s^{(0)} 
= - \overline{\left(\frac{T^\prime}{T^{(0)}}+\frac{\rho^\prime}{\rho^{(0)}}\right)\left( \frac{\partial s^\prime}{\partial t}
+\pmb{v}^\prime\cdot\nabla s^{(0)}+\pmb{v}^{(0)}\cdot\nabla s^\prime\right)}
-\overline{\pmb{v}^\prime\cdot\nabla s^\prime}
 + {\epsilon^{(2)}\over T^{(0)}}+{\rho^{(2)}\epsilon^{(0)}-\overline{\nabla\cdot\pmb{F}^{(2)}}+\overline{\rho^\prime\epsilon^\prime}\over \rho^{(0)}T^{(0)}},
\label{eq:ent2eq}
\end{eqnarray}
\be
\pmb{F}^{(2)}=-\lambda^{(0)}_{\rm rad}\nabla T^{(2)}-\lambda^{(2)}_{\rm rad}\nabla T^{(0)}-\overline{\lambda^\prime_{\rm rad}\nabla T^\prime},
\label{eq:radtrans2}
\ee
where the first order quantities $f^{(1)}$ are simply written as $f'$, and 
$\Phi^{(2)}$ and $\nabla^2\Phi^{(2)}=4\pi G\rho^{(2)}$ are ignored in the Cowling approximation.

For later convenience, we denote the right hand side of equation (\ref{eq:secondmomcon}) as $\pmb{G}^{(2)}$, that is, 
\be
\pmb{G}^{(2)}\equiv -  {\pmb{v} ' \cdot \nabla \pmb{v} '}  + g {\left( {\frac{{\rho '}}{{\rho ^{\left( 0 \right)} }}} \right)^2}  
\pmb{e}_r 
+ \frac{1}{{\rho ^{\left( 0 \right)} }} {\frac{{\rho '}}{{\rho ^{\left( 0 \right)} }}\nabla p'}.
\label{eq:gdef}
\ee
We may write $\pmb{G}^{(2)}$, depending on the basis vector set, as
\be
\pmb{G}^{(2)}=G_r^{(2)}\pmb{e}_r+G_\theta^{(2)}\pmb{e}_\theta+G_\phi^{(2)}\pmb{e}_\phi=
G_r^{(2)}\pmb{e}_r+G_q^{(2)}\pmb{e}_q+G_{\bar q}^{(2)}\pmb{e}_{\bar q},
\ee
where
\be
\pmb{e}_q={\pmb{e}_\theta+\rmi\pmb{e}_\phi\over\sqrt{2}}, \quad
\pmb{e}_{\bar q}={\pmb{e}_\theta-\rmi\pmb{e}_\phi\over\sqrt{2}}.
\ee

\subsection{mean flow equations}

The set of equations derived above are coupled linear partial differential equations for the second order axisymmetric perturbations where
$(r,\theta)$ and $t$ are the independent variables.
The products of first order perturbations provide inhomogeneous terms.
To describe the $\theta$ dependence of the second order perturbations, we use series expansions of finite length, denoted by $k_{\rm max}$, in terms of spherical harmonic functions $Y_l^{0}(\theta,\phi)$.
The velocity perturbation $\pmb{v}^{(2)}$ is expanded as
\be
{v_r^{(2)}}(\pmb{x},t)=\sum_{k=1}^{k_{\rm max}}\hat v_{S,l_k}^{(2)}(r,t)Y_{l_k}^0(\theta,\phi),
\label{eq:expand2r}
\ee
\be
{v_\theta^{(2)}}(\pmb{x},t)
=\sum_{k=2}^{k_{\rm max}}\hat v_{H,l_k}^{(2)}(r,t)\frac{\partial}{\partial\theta}
Y_{l_k}^0(\theta,\phi),
\label{eq:expand2h}
\ee
\be
{v_\phi^{(2)}}(\pmb{x},t)
=-\sum_{k=1}^{k_{\rm max}}  \hat v_{T,l'_k}^{(2)}(r,t)\frac{\partial}{\partial\theta}Y_{l'_k}^0(\theta,\phi),
\label{eq:expand2t}
\ee
and the pressure
perturbation $p^{(2)}$ as
\be
p^{(2)}(\pmb{x},t)=\sum_{k=1}^{k_{\rm max}}p_{l_k}^{(2)}(r,t)Y_{l_k}^0(\theta,\phi),
\label{eq:expand2pr}
\ee
where $l_k=2(k-1)$ and $l^\prime_k=l_k+1$ 
for $k=1,~2,~\cdots, ~k_{\rm max}$.

By substituting the expansions (\ref{eq:expand2r}) to
(\ref{eq:expand2pr}) into equations (\ref{eq:secondmomcon}) to
(\ref{eq:radtrans2}), multiplying by a given spherical harmonic function, 
and integrating over solid angle, we derive a finite set of differential
equations for the expansion coefficients, which depend on $r$ and $t$ (see Lee et al 2016).  
When we integrate over solid angle the non-linear terms
such as $ (Y_{l_k}^0)(\overline{\pmb{v}' \cdot \nabla
\pmb{v}'})_r$, we have to evaluate angular integration
of products of three spherical harmonic functions, and we carry out the integration
by introducing spin-weighted spherical harmonic functions
${}_sY_l^m(\theta,\phi)$ (see, e.g., Newman \& Penrose 1966; Varshalovich et al. 1988).
Since the set of equations are linear equations for the second order expansion coefficients, 
we look for solutions whose time dependence is given by $e^{\gamma t}$.
Replacing the time derivatives $\partial/\partial t$ by $\gamma$, the finite set of partial linear
differential equations reduces to a set of linear ordinary differential equations that possess inhomogeneous terms.

Using vector notation, we formally write the set of linear ordinary differential equations
with inhomogeneous terms as
\be
r{d\pmb{Z}\over dr}=\pmbmt{A}(r,\gamma)\pmb{Z}+\pmb{I}(r,\omega),
\label{eq:mfe}
\ee
where
\be
\pmb{Z}=\left(\matrix{\pmb{z}_1\cr\pmb{z}_2\cr\pmb{z}_3\cr\pmb{z}_4\cr}\right), \quad 
\pmb{z}_1=\left(\matrix{\hat v_{S,{l_1}}^{(2)}/r\sigma_0\cr \hat v_{S,{l_2}}^{(2)}/r\sigma_0\cr \vdots \cr}\right), \quad 
\pmb{z}_2=\left(\matrix{p^{(2)}_{l_1}/gr\rho^{(0)}\cr p^{(2)}_{l_2}/gr\rho^{(0)}\cr \vdots \cr}\right), \quad
\pmb{z}_3=\left(\matrix{L^{(2)}_{r,l_1}/L^{(0)}_r\cr L^{(2)}_{r,l_2}/L^{(0)}_r\cr \vdots \cr}\right), \quad
\pmb{z}_4=\left(\matrix{T^{(2)}_{l_1}/T^{(0)}\cr T^{(2)}_{l_2}/T^{(0)}\cr \vdots \cr}\right),
\ee
and $\pmbmt{A}$ and $\pmb{I}$ respectively represent the coefficient matrix and the inhomogeneous term
(see Lee et al. 2016).

Non-radial components of equation (\ref{eq:secondmomcon}) provide auxiliary equations, given by
\be
\pmbmt{W}\left(\matrix{\pmb{z}_h\cr\pmb{z}_t\cr}\right)=\left(\matrix{0 & \bar f\pmbmt{\Lambda}_0^{1/2}\cr
-\sqrt{2}\bar f\pmbmt{C}_C^0 &0\cr}\right)\left(\matrix{\pmb{z}_1\cr\pmb{z}_2/c_1\cr}\right)
+{1\over \sqrt{2}gc_1}\left(\matrix{\overline{\pmb{G}_q^0-\pmb{G}_{\bar q}^0}\cr\overline{-\rmi (\pmb{G}_q^1+\pmb{G}_{\bar q}^1)}\cr}\right),
\label{eq:aux}
\ee
where 
\be
\pmbmt{W}=\left(\matrix{-\bar\gamma\pmbmt{I}& \bar f\pmbmt{C}_B^1\cr -\bar f\pmbmt{C}_B^0 & -\bar\gamma\pmbmt{I}\cr}\right)
\ee
with $\pmbmt{I}$ being the identity matrix, and 
\be
\pmb{z}_h=\left(\matrix{\sqrt{\Lambda_{l_1}}\hat v_{H,{l_1}}^{(2)}/r\sigma_0\cr \sqrt{\Lambda_{l_2}}\hat v_{H,{l_2}}^{(2)}/r\sigma_0\cr \vdots \cr}\right), \quad 
\pmb{z}_t=\left(\matrix{\sqrt{\Lambda_{l'_1}}\hat v_{T,{l'_1}}^{(2)}/r\sigma_0\cr \sqrt{\Lambda_{l'_2}}\hat v_{T,{l'_2}}^{(2)}/r\sigma_0\cr \vdots \cr}\right),
\ee
where $\bar\gamma=\gamma/\sigma_0$, $\bar f=\sqrt{4\pi/3}\bar\Omega$, and $\Lambda_l=l(l+1)$.
See Lee et al (2016) for the definition of the matrices $\pmbmt{\Lambda}_0^{1/2}$, $\pmbmt{C}_B^0$, $\pmbmt{C}_B^1$,
and $\pmbmt{C}_C^0$.
The vectors $\overline{\pmb{G}_q^j}$ and $\overline{\pmb{G}_{\bar q}^j}$ for $j=0$ and $1$ on the right hand side of equation (\ref{eq:aux}) come from the vector $\pmb{G}^{(2)}$
defined by equation (\ref{eq:gdef}).
The $k$-th components of the vectors $\overline{\pmb{G}_q^0}$, $\overline{\pmb{G}_{\bar q}^0}$, $\overline{\pmb{G}_q^1}$, and 
$\overline{\pmb{G}_{\bar q}^1}$ are given by
\be
(\overline{\pmb{G}_q^0})_k=\int{}_1Y_{l_k}^0\overline{G_q^{(2)}}do, \quad 
(\overline{\pmb{G}_{\bar q}^0})_k=\int{}_{-1}Y_{l_k}^0\overline{G_{\bar q}^{(2)}}do, \quad
(\overline{\pmb{G}_q^1})_k=\int{}_1Y_{l'_k}^0\overline{G_q^{(2)}}do, \quad 
(\overline{\pmb{G}_{\bar q}^1})_k=\int{}_{-1}Y_{l'_k}^0\overline{G_{\bar q}^{(2)}}do,
\ee
where $do=\sin\theta d\theta d\phi$, and $l_k=2k-2$ and $l'_k=l_k+1$ for $k=1,~2,~3,~\cdots$.
We note that
$
\pmbmt{W}=-\pmbmt{W}^T
$
and $\pmbmt{W}$ becomes singular when $\bar\gamma=0$.
For non-zero values of $\bar\gamma$, 
we eliminate the variables $\pmb{z}_h$ and $\pmb{z}_t$ to derive equation (\ref{eq:mfe})
from the set of perturbed equations of second order, that is, we have to invert
the matrix $\pmbmt{W}$ for the elimination, which becomes numerically difficult when $|\bar\gamma|$ is extremely small.

The $\bar\gamma$ value may be determined by various processes.
Dissipative effects such as non-adiabatic one in binary stars affect through tidal interactions the binary evolution which is described by slow changes of the binary parameters.
The magnitude of the tidal effects on the primary is given by the parameter $f_0\propto qa_{\rm orb}^{-3}$, 
which depends on $a_{\rm orb}$ for a given value of $q$.
The change rate of the mean separation $a_{\rm orb}$ may be given by 
(e.g., Savonije \& Papaloizou 1997; Witte \& Savonije 2002; Ogilvie \& Lin 2004)
\be
{1\over a_{\rm orb}}{da_{\rm orb}\over dt}={1\over |E_{\rm orb}|}{dE_{\rm orb}\over dt}=
{1\over |E_{\rm orb}|}{n\Omega_{\rm orb}\over m}{\cal T}_2=-{1\over |E_{\rm orb}|}{\Omega_{\rm orb}}{\cal T}_2,
\ee
where ${\cal T}_2$ is defined by equation (\ref{eq:tidaltorque}), $E_{\rm orb}=-{GMM_2/ 2a_{\rm orb}}$,
and we write the tidal forcing frequency as $\omega_{\rm tide}=n\Omega_{\rm orb}+m\Omega$
with $n=-m=2$.
The normalized growth (decay) rate $\bar\gamma_{\rm tide}\sim 2d\ln f_0/d(\sigma_0t)=-6d\ln a_{\rm orb}/d(\sigma_0t)$ of the linear tidal responses may be given by
\be
\bar\gamma_{\rm tide}=-6{d\ln a_{\rm orb}/ d(\sigma_0t)}
=6q^{-1}{(1+q)^{1/2}}\left({a_{\rm orb}/ R}\right)^{-1/2}\overline{{\cal T}_2},
\label{eq:gammatide}
\ee
where the sign of $\bar\gamma_{\rm tide}$ coincides with that of $\overline{{\cal T}_2}$.
For $M_2\sim 0.1M$ and $a_{\rm orb} \sim 10R~{\rm to}~ 10^2R$, for example, we have $d\ln a_{\rm orb}/d(\sigma_0 t)\sim -\overline{{\cal T}_2}$ and $f_0\sim 10^{-4}~{\rm to}~ 10^{-7}$.
For this parameter range, even at resonance with $g$-modes,
the amplitudes of the tidal responses $S_l$, which is proportional to $f_0$, 
is much smaller than unity (see Figure \ref{fig:tidalresponses}), and
the magnitude of the normalized tidal torque $\overline{{\cal T}_2}$, which is proportional to $f_0^2$, 
is at most of order of $10^{-10}$ or smaller (see Figure \ref{fig:t2}).
This suggests that $\bar\gamma_{\rm tide}$ is in general much smaller than $|2\bar\omega_{\rm I}|$ given in Table 1.
If the orbital shrinkage takes place due to gravitational wave radiation, we may have (see, e.g., Landau \& Lifshitz 1975)
\begin{eqnarray}
\bar\gamma_{\rm grav}=-6{d\ln a_{\rm orb}/ d(\sigma_0 t)}=({48\sqrt{2}/ 5})q(1+q)(R_g/R)^{5/2}(a_{\rm orb}/R)^{-4},
\end{eqnarray}
where $R_g=2GM/c^2$, and $\bar\gamma_{\rm grab}$ is generally smaller than $\bar\gamma_{\rm tide}$.

In this paper, 
we treat $\bar\gamma$ as a constant parameter of order of
$10^{-8}$ so that we can properly inverse the matrix $\pmbmt{W}$.
We confirm that the flow patterns of tidally driven mean flows for $|\bar\gamma|=10^{-8}$ are the same as those for
$|\bar\gamma|=10^{-10}$, and that for sufficiently small values of $\bar\gamma$,
the magnitudes of $v_\phi^{(2)}$ scales as $\gamma v_\phi^{(2)}\approx
{\rm constant}$ where the constant is almost independent of $\bar\omega_{\rm tide}$.

\subsection{angular momentum transport by waves}

The angular momentum transport by waves in rotating stars may be described by (e.g., Lee 2013, see also Grimshaw 1984)
\be
\tilde \rho{d\over dt}\overline{\ell\left(\pmb{x}+\pmb{\xi}\right)}=-\nabla\cdot\overline{\left(\pmb{\xi}{\partial p'\over\partial\phi}\right)}
-\nabla\cdot\overline{\left(\rho\pmb{\xi}{\partial\Phi'\over\partial\phi}\right)}-\overline{\rho'{\partial\Phi'\over\partial\phi}},
\label{eq:angtransp}
\ee
where $\pmb{\xi}$ is the displacement vector associated with the wave, and 
\begin{eqnarray}
\overline{\ell(\pmb{x}+\pmb{\xi})}=\overline{\left[\left(\pmb{x}+\pmb{\xi}\right)\times\pmb{v}\left(\pmb{x}+\pmb{\xi}\right)\right]}\cdot\pmb{e}_z\equiv \ell^{(0)}+\ell^{(2)}
\end{eqnarray}
is the specific angular momentum in the $z$-direction, where
\be
\ell^{(0)}=(r\sin\theta)^2\Omega,
\ee
\be
\ell^{(2)}=r\sin\theta v_\phi^{(2)}+r\sin\theta\overline{v_{\phi;j}^{(1)}\xi_j}
+\overline{\left(v_\phi^{(1)}\xi_r-\xi_\phi v_r^{(1)}\right)\sin\theta+\left(v_\phi^{(1)}\xi_\theta-\xi_\phi v_\theta^{(1)}\right)\cos\theta}
+\left[\overline{\left(\xi_r\sin\theta+\xi_\theta\cos\theta\right)^2+\xi_\phi^2}\right]\Omega.
\ee
The total time derivative on the left-hand-side of equation (\ref{eq:angtransp}) is defined as
\be
{d\over dt}={\partial\over\partial t}+\overline{v_r(\pmb{x}+\pmb{\xi})}{\partial\over\partial r}
+\overline{v_\theta(\pmb{x}+\pmb{\xi})}{1\over r}{\partial\over\partial \theta}
+\overline{v_\phi(\pmb{x}+\pmb{\xi})}{1\over r\sin\theta}{\partial\over\partial \phi},
\ee
where
\be
\overline{v_i(\pmb{x}+\pmb{\xi})}= v_i^{(0)}(\pmb{x})+\delta v_i^{(2)}(\pmb{x}),
\ee
and
\be
\delta v_i^{(2)}(\pmb{x})=v_i^{(2)}(\pmb{x})+\overline{v_{i;j}^{(1)}(\pmb{x})\xi_j(\pmb{x})}
+{1\over 2}v_{i;j;k}^{(0)}(\pmb{x})\overline{\xi_j(\pmb{x})\xi_k(\pmb{x})}
\ee
is the second order Lagrangian perturbation of the velocity, and the semicolon indicates covariant derivatives.
Note that for uniform rotation, $v^{(0)}_{i;j;k}=0$ and hence 
$\delta v_i^{(2)}(\pmb{x})=v_i^{(2)}(\pmb{x})+\overline{v_{i;j}^{(1)}(\pmb{x})\xi_j(\pmb{x})}$.
The treatment given above is based on the Lagrangian mean wave-mean flow interaction theory developed by
Andrews \& McIntyre (1976, 1978ab).
See also Dunkerton (1980), Grimshaw (1984), and B\"uhler (2014) for reviews of wave-mean flow interaction theories.

Integrating equation (\ref{eq:angtransp}) over the whole volume of the star, we obtain
\be
\int dV\tilde\rho{d\over dt}\overline{\ell(\pmb{x}+\pmb{\xi})}=-\int dV\overline{\rho'{\partial\Phi'\over \partial\phi}},
\ee
where $dV=d^3\pmb{x}$ and $\tilde\rho(\pmb{x}) d^3\pmb{x}=\rho(\hat{\pmb{x}})d^3\hat{\pmb{x}}$ with $\hat{\pmb{x}}=\pmb{x}+\pmb{\xi}(\pmb{x})$, and we have ignored the surface term assuming $p'/\rho$ is finite at the surface, that is, $p'/\rho$ remains finite as $\rho\rightarrow 0$
toward the surface.
The right-hand-side becomes the tidal torque when we replace $\Phi'$ by the tidal potential $\Phi'_e+\Psi$.

If we assume $v_r^{(0)}=v_\theta^{(0)}=0$ for uniform rotation, equation (\ref{eq:angtransp}) becomes
\be
\rho\left[{\partial\over\partial t}\ell^{(2)}+\left(\delta v_r^{(2)}{\partial\over\partial r}+\delta v_\theta^{(2)}{1\over r}{\partial\over\partial\theta}\right)\ell^{(0)}\right]=-\nabla\cdot\overline{\left[\rho\pmb{\xi}{\partial \over\partial\phi}\left({p'\over\rho}+\Phi'\right)\right]}
-\overline{{\rho'}{\partial\Phi'\over\partial\phi}},
\label{eq:angtrans}
\ee
where $\Phi'=\Phi'_e+\Psi\propto Y_{l=2}^{m=-2}$ for tidal responses.
Integrating over solid angle, we obtain
\be
\int do \left[{\partial\over\partial t}\ell^{(2)}+\left(\delta v_r^{(2)}{\partial\over\partial r}+\delta v_\theta^{(2)}{1\over r}{\partial\over\partial\theta}\right)\ell^{(0)}\right]={m\over 2\rho r^2}{\partial\over\partial r}r^2{\rm Im}\left[\rho\int \xi_r^*\left({p'\over\rho}+\Phi'\right)do\right]
+{m\over 2}{\rm Im}\left(\int {\rho'\over \rho}^*{\Phi'}do\right),
\label{eq:defw}
\ee
and it is convenient to denote the right hand side of equation (\ref{eq:defw}) as ${\cal W}(r)$, that is,
\be
{\cal W}(r)\equiv {m\over 2\rho r^2}{\partial\over\partial r}r^2{\rm Im}\left[\rho\int \xi_r^*\left({p'\over\rho}+\Phi'\right)do\right]
+{m\over 2}{\rm Im}\left(\int {\rho'\over \rho}^*{\Phi'}do\right).
\label{eq:defw22}
\ee

Equation (\ref{eq:angtransp}) may be regarded as a mean flow equation that describes responses of the
mean flow to waves.
If the waves are non-dissipative, the right-hand-side of equation (\ref{eq:angtransp}) vanishes, 
indicating conservation of the specific angular momentum $\overline{ \ell(\hat{\pmb{x}}) }$
as stated by Goldreich \& Nicholson (1989).
Responses of the waves to mean flows, on the other hand, may be described by the equation for wave action.
In the Lagrangian mean theory (Andrwes \& McIntyre 1978ab; see also Dunkerton 1980; Grimshaw 1984), 
the wave action $A$ obeys
\be
{dA\over dt}+\tilde\rho^{-1}\nabla\cdot\pmb{B}=D,
\label{eq:waveaction}
\ee
where
\be
A=\overline{\sum_i(\partial\xi_{i}/\partial\phi)(\pmb{v}^l+\pmb{\Omega}\times\pmb{\xi})_i}, \quad B_j=\overline{p(\hat{\pmb{x}})\sum_i(\partial\xi_{i}/\partial\phi)K_{ij}}, \quad \pmb{v}^l=\pmb{v}(\hat{\pmb{x}})-\overline{\pmb{v}(\hat{\pmb{x}})},
\ee
and $K_{ij}$ is the $(i,j)$th cofactor of the Jacobian $J\equiv {\rm det}(\partial\hat{\pmb{x}}/\partial\pmb{x})$,
and in small amplitude limit of the waves the dissipation term $D$ reduces to
\be
 D=-\rho^{-1}{\alpha_T}\overline{{\partial\delta p\over\partial \phi}{\delta s\over c_p}}.
\ee
It is dissipative processes that cause interaction between the mean flows and waves.

\subsection{velocity fields of tidally driven mean flows}

For tidally driven mean flows, we calculate the velocity fields $\pmb{v}_H^{(2)}=v_y^{(2)}\pmb{e}_y+v_z^{(2)}\pmb{e}_z$ on
three spherical surfaces of different radii $r/R=0.99$, 0.95 and 0.90, where, assuming the $x$-axis is towards 
the observer, the velocity fields $(v_y,v_z)$ in the $y$-$z$ plane are given by
\be
v_y^{(2)}=v_r^{(2)}\sin\theta\sin\phi+v_\theta^{(2)}\cos\theta\sin\phi+v_\phi^{(2)}\cos\phi,
\ee
\be
v_z^{(2)}=v_r^{(2)}\cos\theta-v_\theta^{(2)}\sin\theta,
\ee
and $\theta$ and $\phi$ are respectively the colatitude, measured from the $z$-axis, and
the azimuthal angle, measured from the $x$-axis.
For the mean flow calculations we use the expansion length $k_{\rm max}=16$, which is we find long enough to get
good convergence of the expansions for the perturbations.
Figure \ref{fig:m15progreso} shows $\pmb{v}_H^{(2)}$ at the forcing frequency $\bar\omega_{\rm tide}=0.3704823794$ that is
in resonance with the prograde $l=-m=2$ $g_6$-mode of the $15M_\odot$ model for $\bar\gamma=-10^{-8}$.
On each of the spherical surfaces, $\pmb{v}_H^{(2)}$ is normalized by its maximum value $v_{\rm max}^{(2)}(x=r/R)\equiv{\rm max}(|\pmb{v}_H^{(2)}(r,\theta,\phi)|)$ on that surface, and 
the length of the arrows is proportional to the magnitude of normalized $\pmb{v}_H^{(2)}$.
As discussed in Lee et al (2016), the $\phi$ component of $\pmb{v}_H^{(2)}$
dominates the $r$ and $\theta$ components, and hence the velocity fields $\pmb{v}_H^{(2)}$ of tidally driven mean flows
are almost parallel to the equator of the star.
The velocity field $\pmb{v}^{(2)}_H$ is symmetric about the equator, and the amplitudes of $\pmb{v}_H^{(2)}$ tend to be confined to the equatorial regions.
Since mean flows arise from non-adiabatic effects accompanied with the responses, 
$v_{\rm max}^{(2)}(x)$ becomes largest in the outer most layers where the non-adiabatic effects become most significant.
For example, we find $v_{\rm max}^{(2)}(x=0.9)\sim 0.01\times v_{\rm max}^{(2)}(x=0.99)$.
In the equatorial regions, the velocities are prograde in the surface layers, while they become retrograde in 
the deep interior, suggesting that there arises differential rotation in radial direction.
The amplitude confinement of $\pmb{v}_H^{(2)}$ into the equatorial regions also indicates 
differential rotation in the $\theta$-direction.

\begin{figure}
\begin{center}
\resizebox{0.33\columnwidth}{!}{
\includegraphics{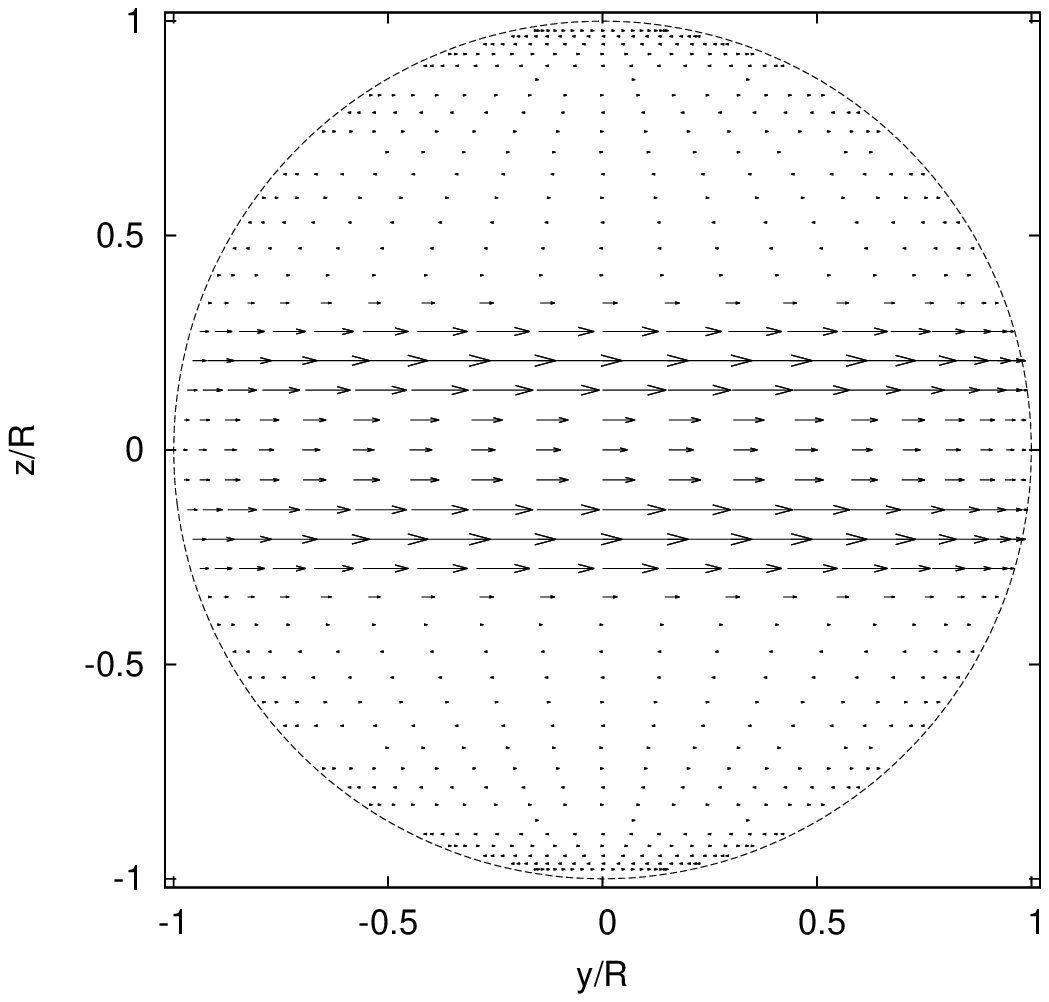}}
\resizebox{0.33\columnwidth}{!}{
\includegraphics{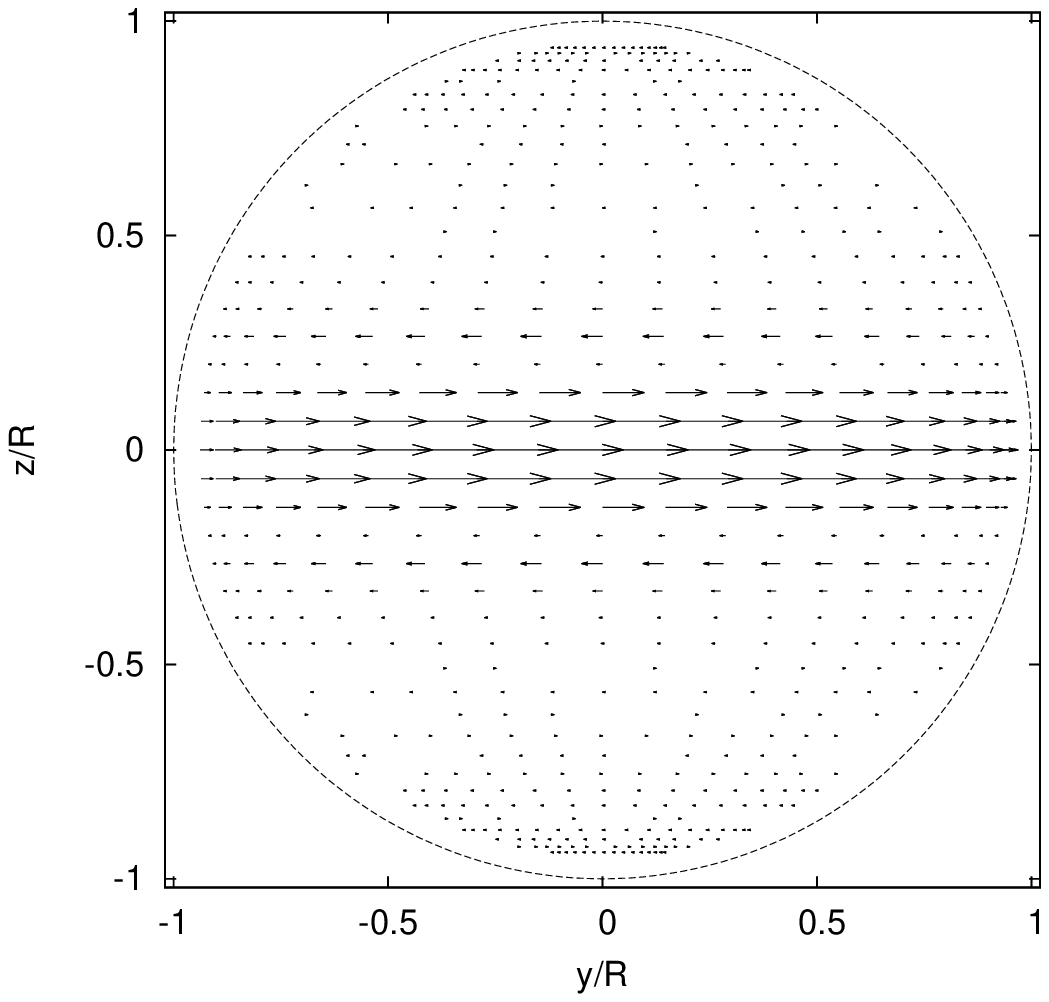}}
\resizebox{0.33\columnwidth}{!}{
\includegraphics{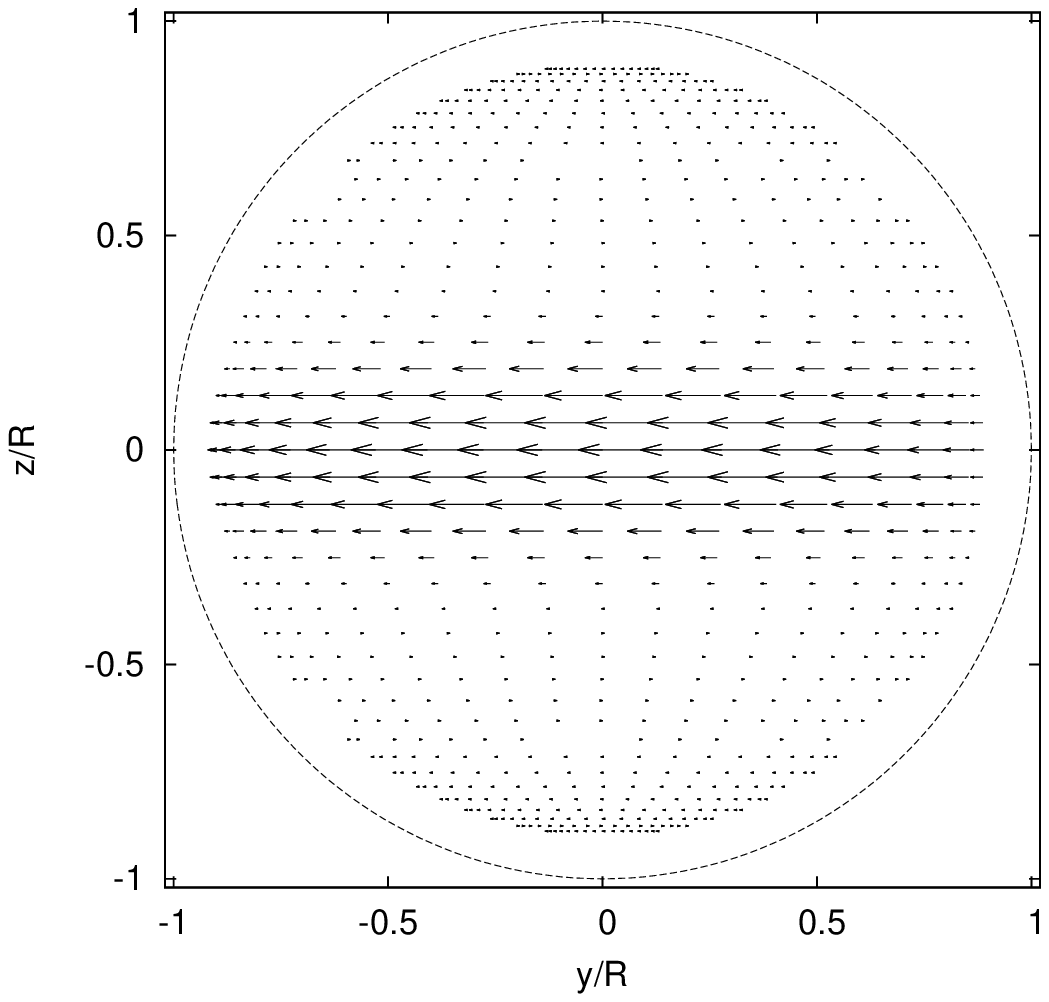}}
\end{center}
\caption{Velocity field $\pmb{v}_H^{(2)}=v_y^{(2)}\pmb{e}_y+v_z^{(2)}\pmb{e}_z$ of the mean flows at  
the forcing frequency $\bar\omega_{\rm tide}=0.3704823794$ in resonance with
the prograde $l=-m=2$ $g_6$-mode for the $15M_\odot$ model, where $\bar\Omega=0.1$ and $\bar\gamma=-10^{-8}$
are assumed.
From left to right panels, $\pmb{v}_H^{(2)}$ on the spherical surfaces of radii $x=r/R=0.99$, 0.95, and 0.90 is plotted.
The length of the arrows is proportional to the magnitude of $\pmb{v}_H^{(2)}$, which is 
normalized by its maximum value 
$v_{\rm max}^{(2)}(x)$ on each of the surfaces.
The ratio of the maximum velocity $v_{\rm max}^{(2)}(x)$ to that on the surface of $x=0.99$ is 0.27 for $x=0.95$ and 0.036 for $x=0.90$, respectively.
}
\label{fig:m15progreso}
\end{figure}

Figure \ref{fig:m15progreso-gamma} shows the velocity fields $\pmb{v}_H^{(2)}$ at the same forcing frequency $\omega_{\rm tide}$ but for $\bar\gamma=+10^{-8}$.
The directions of the velocity $\pmb{v}_H^{(2)}\approx v_y^{(2)}\pmb{e}_y$ are opposite to those for $\bar\gamma=- 10^{-8}$, but we find that the directions of $v_x^{(2)}$ and $v_z^{(2)}$ remain the same.
As indicated by equation (\ref{eq:angtrans}), only the term $\partial\ell^{(2)}/\partial t=\gamma\ell^{(2)}$ explicitly 
depend on $\gamma$, and the terms other than $v_\phi^{(2)}$ in $\ell^{(2)}$ are products of the first order perturvations, which are assumed to be proportional to $e^{\gamma t/2}$.
If $\partial v_\phi^{(2)}/\partial t=\gamma v_\phi^{(2)}$ is dominating, $v_\phi^{(2)}$ at a given $r$ has to change its sign according to the sign of $\gamma$ to balance the right-hand-side of equation (\ref{eq:angtrans}), which does not explicitly depend on $\gamma$.

Figure \ref{fig:g_offreso_prog} shows $\pmb{v}_H^{(2)}$ of mean flows at a forcing frequency $\omega_{\rm tide}$
in off-resonance with low frequency modes of the star, where we use $\bar\omega_{\rm tide}=0.35$ for
$\bar\gamma=-10^{-8}$.
The magnitudes of $\pmb{v}_H^{(2)}$ are much smaller than those at resonance with the $g_6$-mode.
This is of course because
the amplitudes of tidal responses in off-resonance are much smaller than those in the resonance.
The velocity fields $\pmb{v}_H^{(2)}$ are confined to equatorial regions in the surface layers, but
the confinement is not necessarily strong in the deep interior, where $\pmb{v}_H^{(2)}$ shows more complicated
behavior as a function of $\theta$.
The velocities $\pmb{v}_H^{(2)}$ are retrograde in the surface layers, but $\pmb{v}_H^{(2)}$ at the equator
can be prograde at the surface.
It may be important to note that the averaged velocities $\int\sin\theta \pmb{v}_H^{(2)}do$ in the deep interior are prograde, 
which may be consistent with the belief that prograde tidal forcing causes
acceleration of rotation rate of the star.

\begin{figure}
\begin{center}
\resizebox{0.33\columnwidth}{!}{
\includegraphics{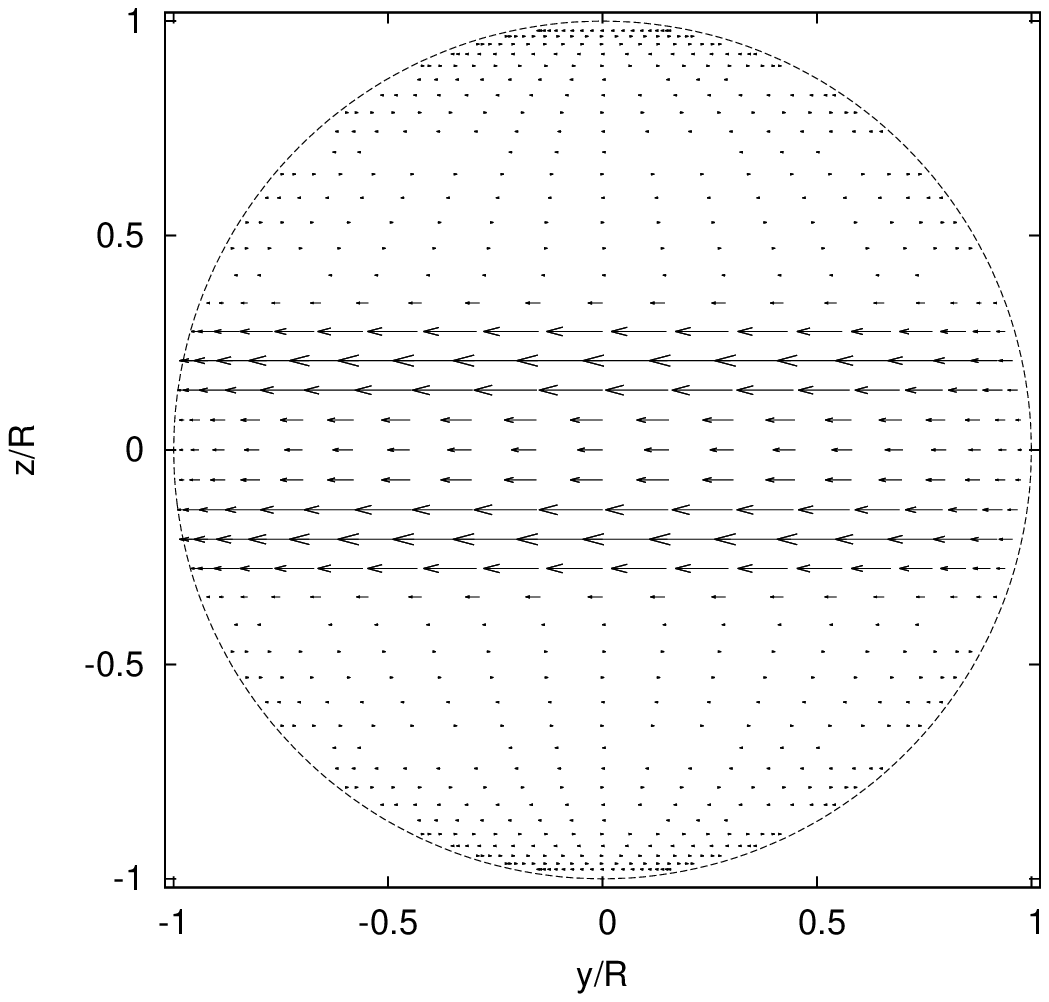}}
\resizebox{0.33\columnwidth}{!}{
\includegraphics{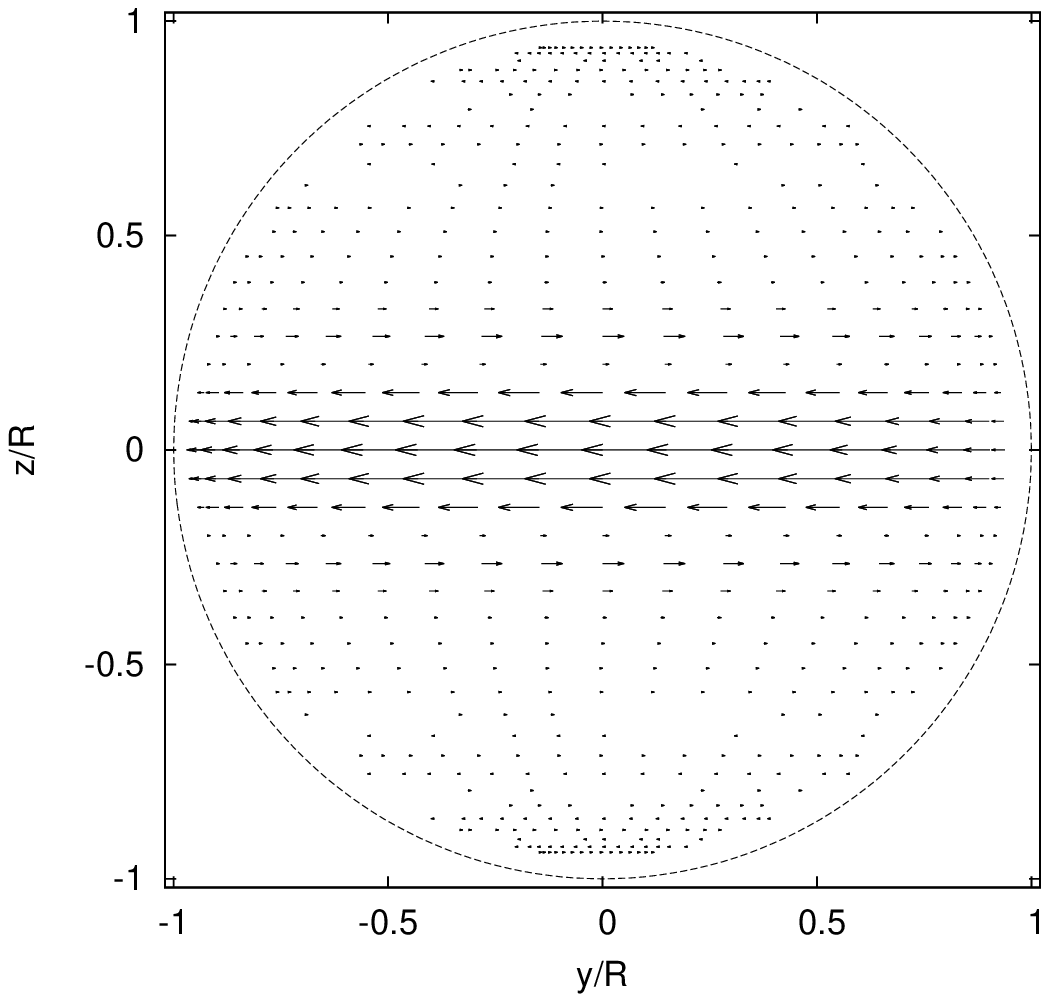}}
\resizebox{0.33\columnwidth}{!}{
\includegraphics{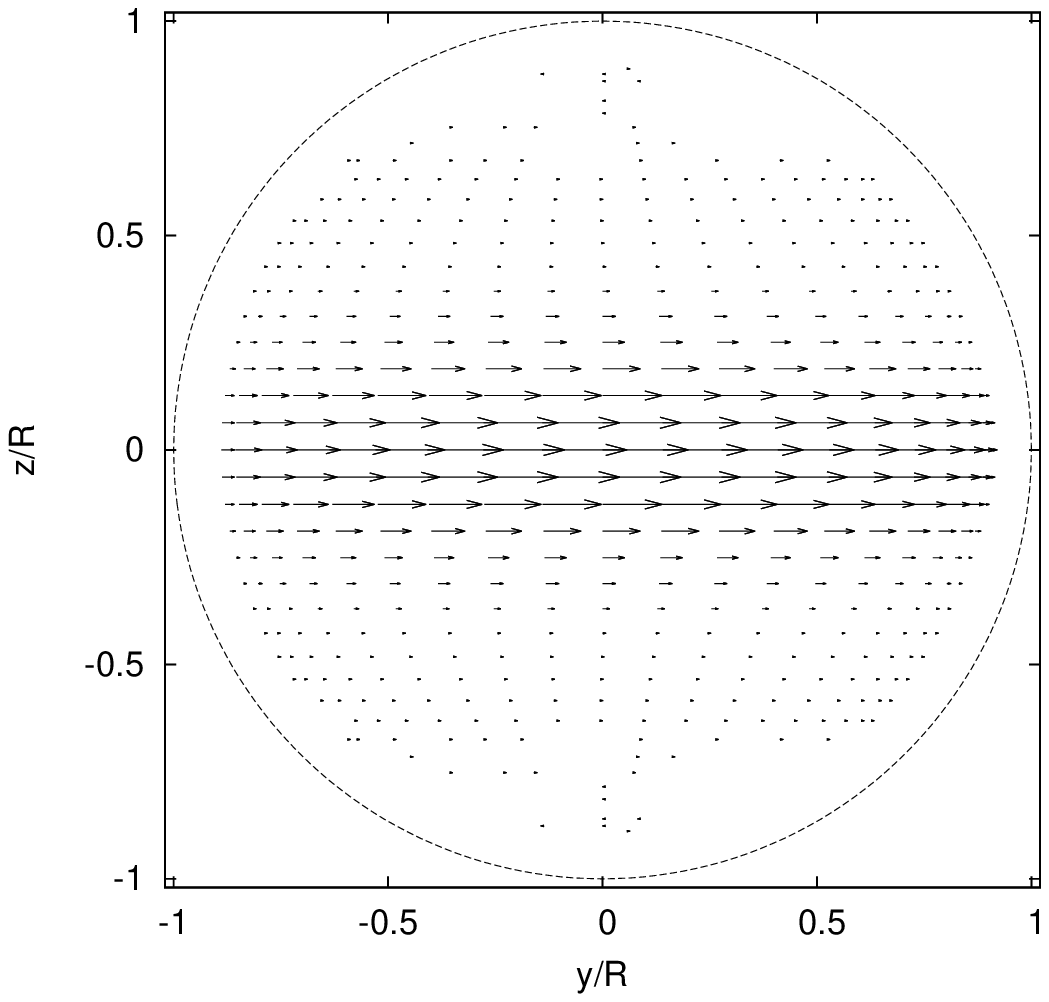}}
\end{center}
\caption{Same as Figure \ref{fig:m15progreso} but for $\bar\gamma=10^{-8}$, where the ratio of the maximum velocity $v_{\rm max}^{(2)}(x)$ to that on the surface of $x=0.99$ is 0.27 for $x=0.95$ and 0.036 for $x=0.90$, respectively.
}
\label{fig:m15progreso-gamma}
\end{figure}

\begin{figure}
\begin{center}
\resizebox{0.33\columnwidth}{!}{
\includegraphics{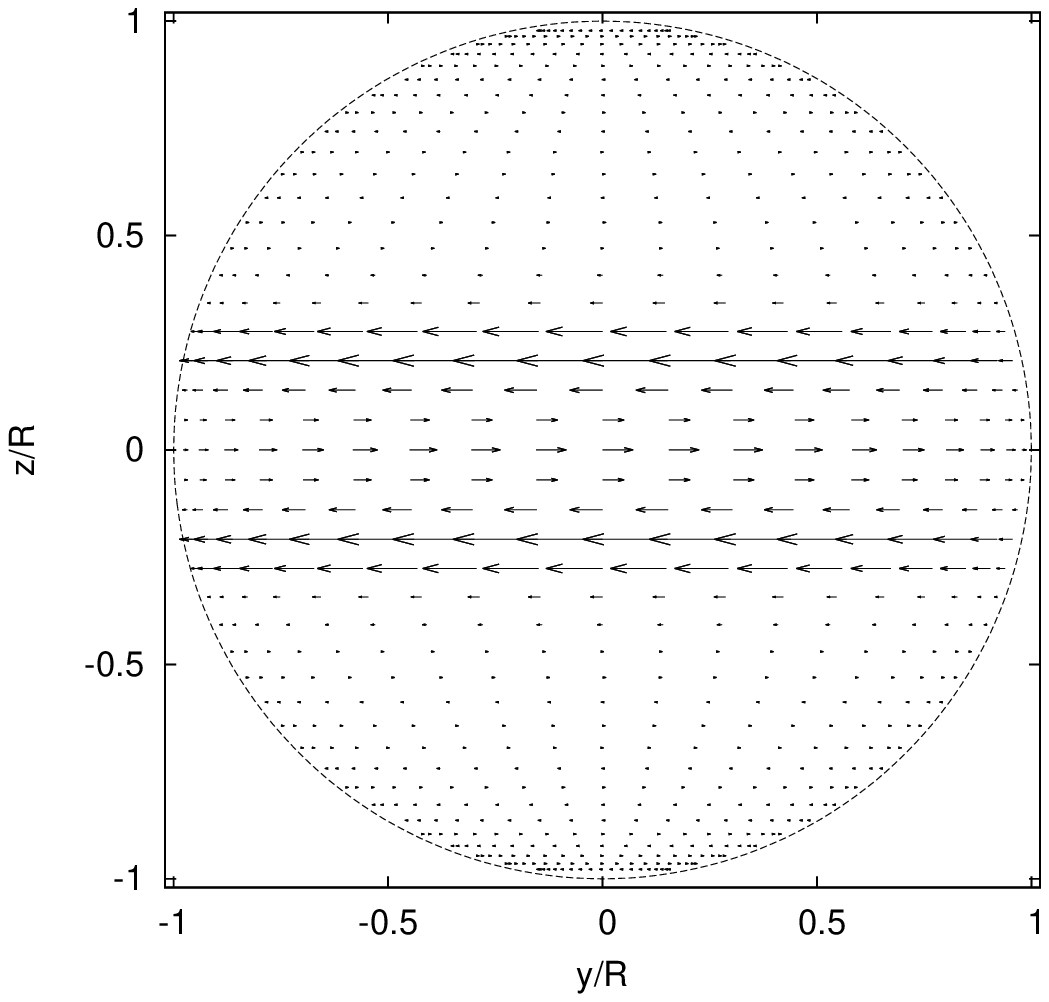}}
\resizebox{0.33\columnwidth}{!}{
\includegraphics{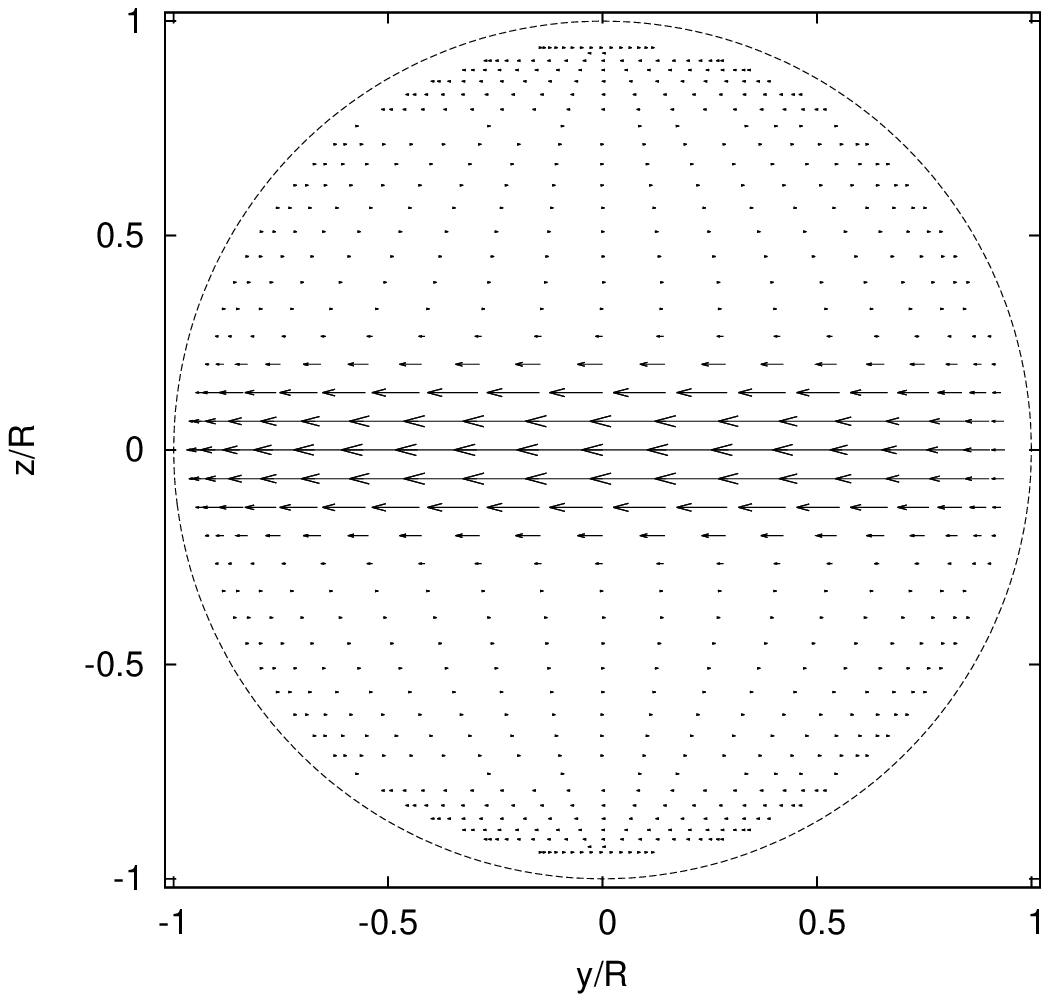}}
\resizebox{0.33\columnwidth}{!}{
\includegraphics{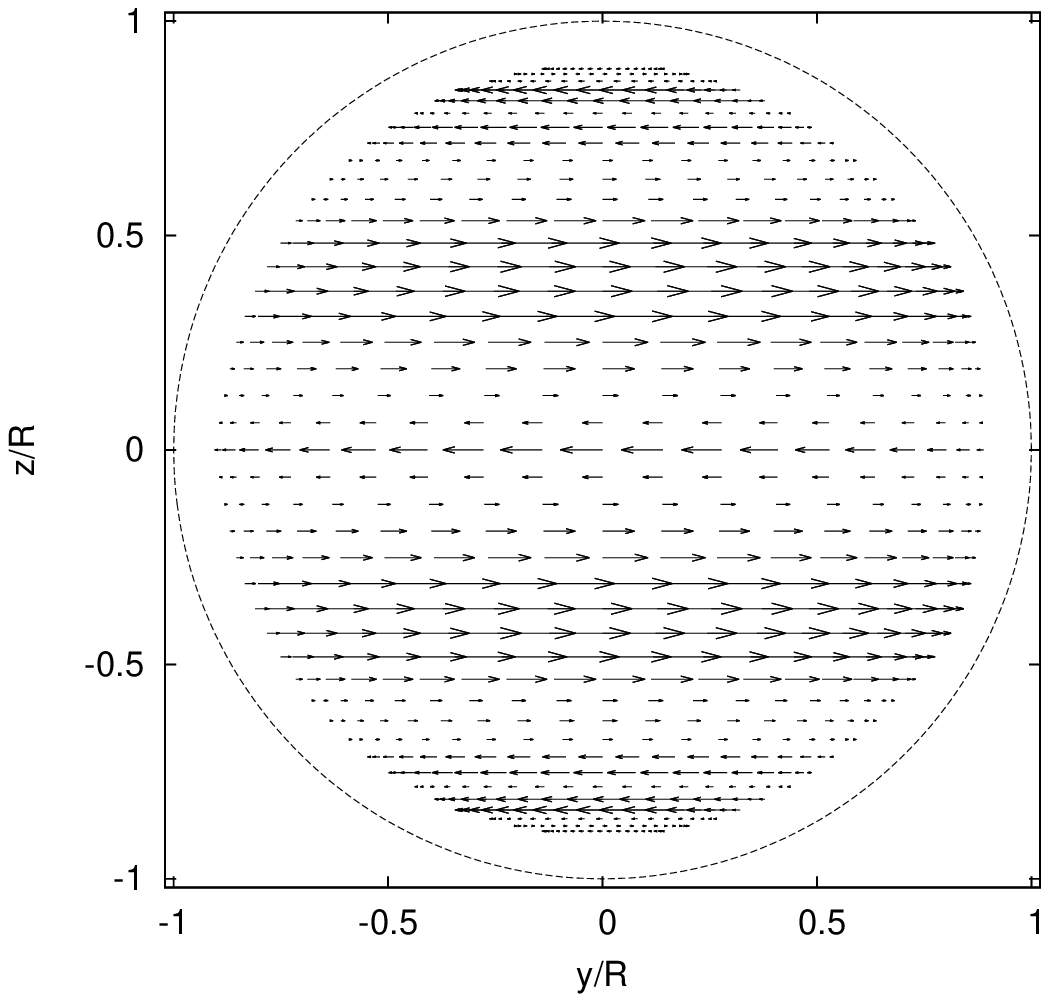}}
\end{center}
\caption{Same as Figure \ref{fig:m15progreso} but for $\bar\omega_{\rm tide}=0.35$, where the ratio of the maximum velocity $v^{(2)}_{\rm max}(x)$ to that on the surface of $x=0.99$ is 0.22 for $x=0.95$ and 0.007 for $x=0.90$, respectively.
}
\label{fig:g_offreso_prog}
\end{figure}

Figure \ref{fig:w-function_prog} plots the function ${\cal W}(x)$ (solid line) at the two tidal forcing frequencies $\bar\omega_{\rm tide}=0.3704823794$
and $\bar\omega_{\rm tide}=0.35$, where the dotted and dashed lines represent the
first and second terms on the right-hand-side of equation (\ref{eq:defw22}), respectively.
The two terms cancel each other to lead to small amplitude ${\cal W}$ in the deep interior.
The function ${\cal W}$ has large amplitudes only in the outer layers of the envelope.
The amplitudes $|{\cal W}|$ at the resonance is much larger than those in off-resonance.
The $r$-dependence of $\cal W$ for the response in the $g$-mode resonance looks quite similar to 
that of the function $(\rho/\rho_m)^{-1}dw/dr$ for the eigen $g$-mode.
This similarity may suggest that in the case of resonant forcing
the velocity fields $v_\phi^{(2)}$ of mean flows are 
closely related to the damping and driving regions for the oscillation mode.
It is interesting to note that the function $\cal W$ for the off-resonance forcing behaves quite differently
from that for the resonant forcing.
The function $\cal W$ for off-resonance forcing is positive in the surface layers, while it is negative for
the resonant forcing.
If the term $\int do~ \partial\ell^{(2)}/\partial t$ is dominating on the left-hand-side of equation (\ref{eq:defw})
and the approximation $\int do~ \partial\ell^{(2)}/\partial t\approx  \int do ~r\sin\theta \partial v_\phi^{(2)}/\partial t$ is valid, 
$\int do~\sin\theta \partial v_\phi^{(2)}/\partial t=\int do~\sin\theta\gamma v_\phi^{(2)}$ is positive (negative) where ${\cal W}$ is positive (negative), which is what we find for the prograde forcing.

\begin{figure}
\begin{center}
\resizebox{0.4\columnwidth}{!}{
\includegraphics{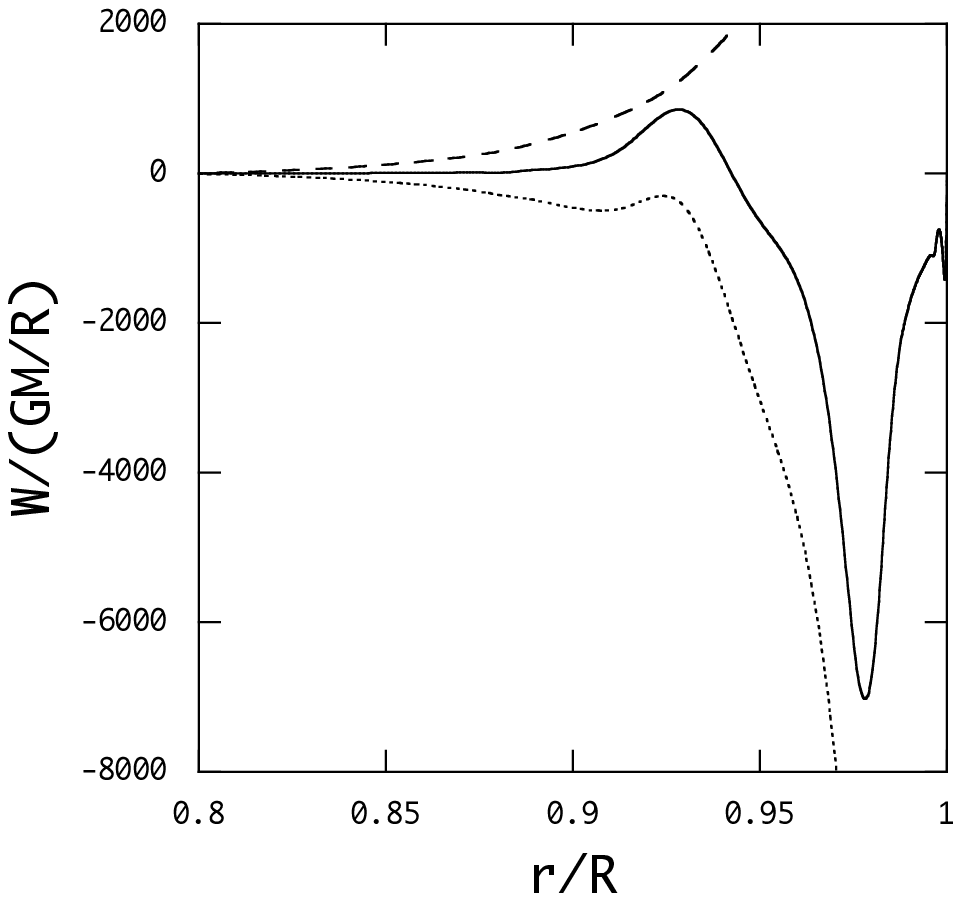}}
\resizebox{0.4\columnwidth}{!}{
\includegraphics{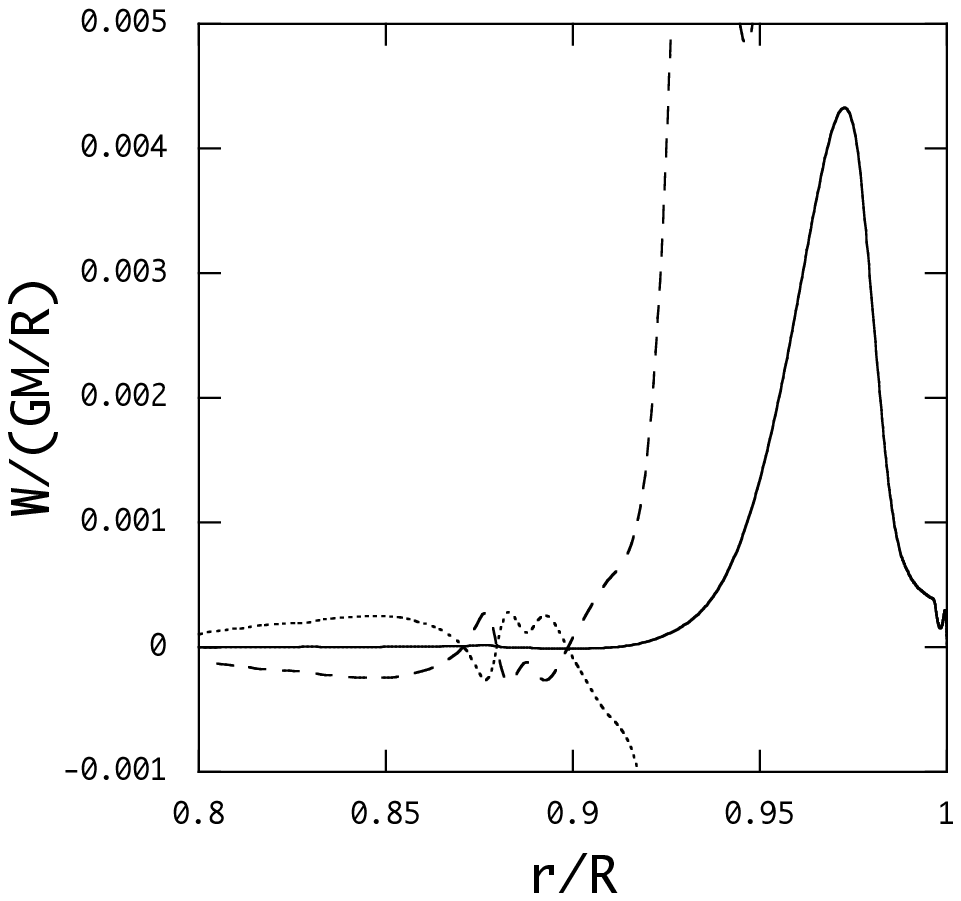}}
\end{center}
\caption{Function ${\cal W}$ (solid line) versus $x=r/R$ for the tidal responses at $\bar\omega_{\rm tide}=0.3704823794$ (left panel) 
and at $\bar\omega_{\rm tide}=0.35$ (right panel) for the $15M_\odot$ model, where $\bar\Omega=0.1$ and $f_0=1$ assumed.
The dotted and dashed lines indicate the first and second terms on the right-hand-side of equation (\ref{eq:defw22}), respectively.
}
\label{fig:w-function_prog}
\end{figure}

Assuming $\bar\gamma=-10^{-8}$, we calculate the mean flow velocity $\pmb{v}_H^{(2)}$ for the retrograde forcing $\omega_{\rm tide}$ in resonance with the $l=-m=2$ $g_7$-mode (Figure \ref{fig:g7reso_retro}) and in off-resonance with $g$-modes (Figure \ref{fig:g_offreso_retro}).
The amplitudes of $\pmb{v}_H^{(2)}$ tend to be confined to the equatorial regions in the surface layers
although this confinement becomes weaker in the deep interior, particularly for the off-resonant forcing.
For the resonant forcing, the velocities $\pmb{v}_H^{(2)}$ are retrograde at the surface but become prograde as
we go into the deep interior, which is consistent with the behavior of the function ${\cal W}(r)$ 
in the left panel of Figure \ref{fig:wfunctions_retro}.
Note that the signs of the function $\cal W$ at a given radial distance $r$ are in general opposite to each other between the prograde and retrograde forcing with similar $|\bar\omega_{\rm tide}|$.
For the off-resonant forcing, 
the velocities $\pmb{v}_H^{(2)}$ are mostly retrograde, which, as shown in the right panel of Figure \ref{fig:wfunctions_retro}, is not consistent with the interpretation in terms of
${\cal W}(r)$ based on the assumption that $\int do~\sin\theta\gamma v_\phi^{(2)}$ is dominating.
This may suggest that the term $\int do~\sin\theta\partial v_\phi^{(2)}/\partial t$ is not necessarily dominating on the left hand side of
equation (\ref{eq:defw}) for off-resonance tidal forcing.

\begin{figure}
\begin{center}
\resizebox{0.33\columnwidth}{!}{
\includegraphics{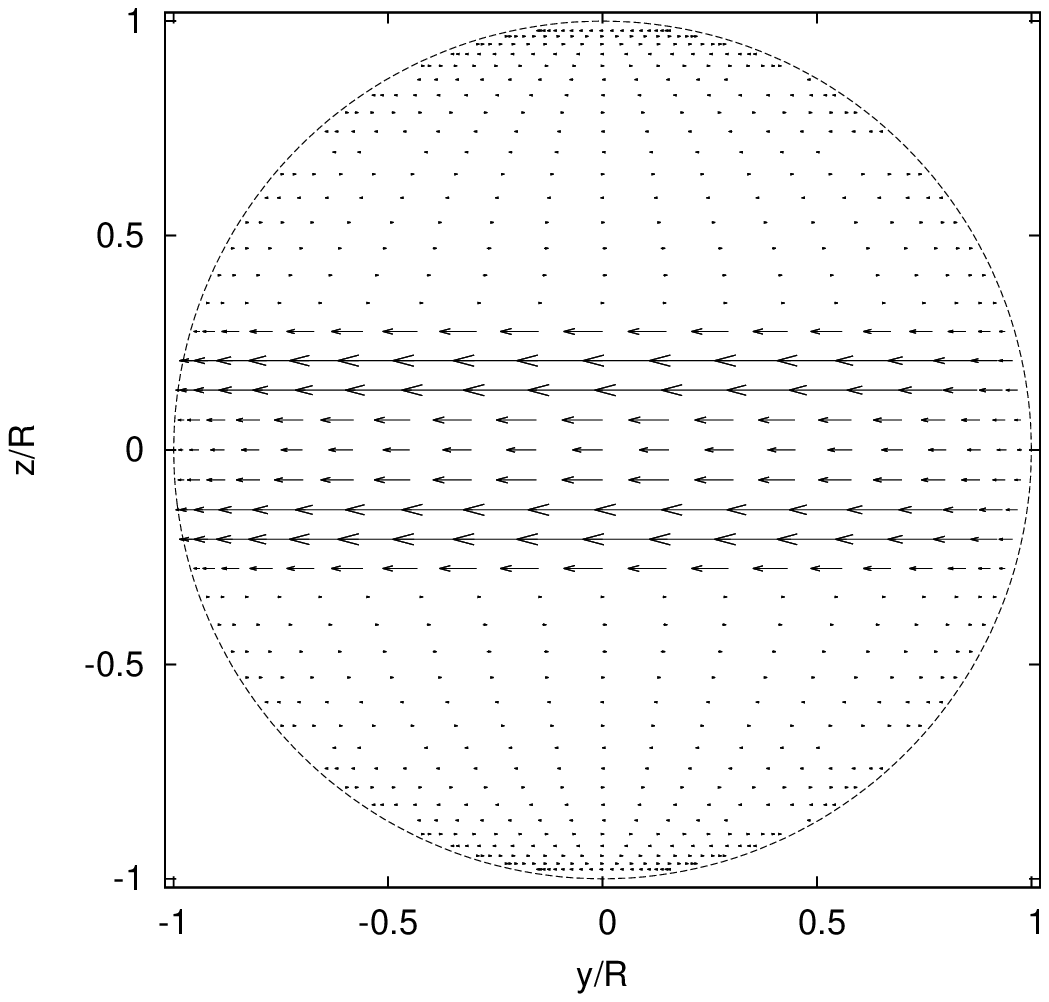}}
\resizebox{0.33\columnwidth}{!}{
\includegraphics{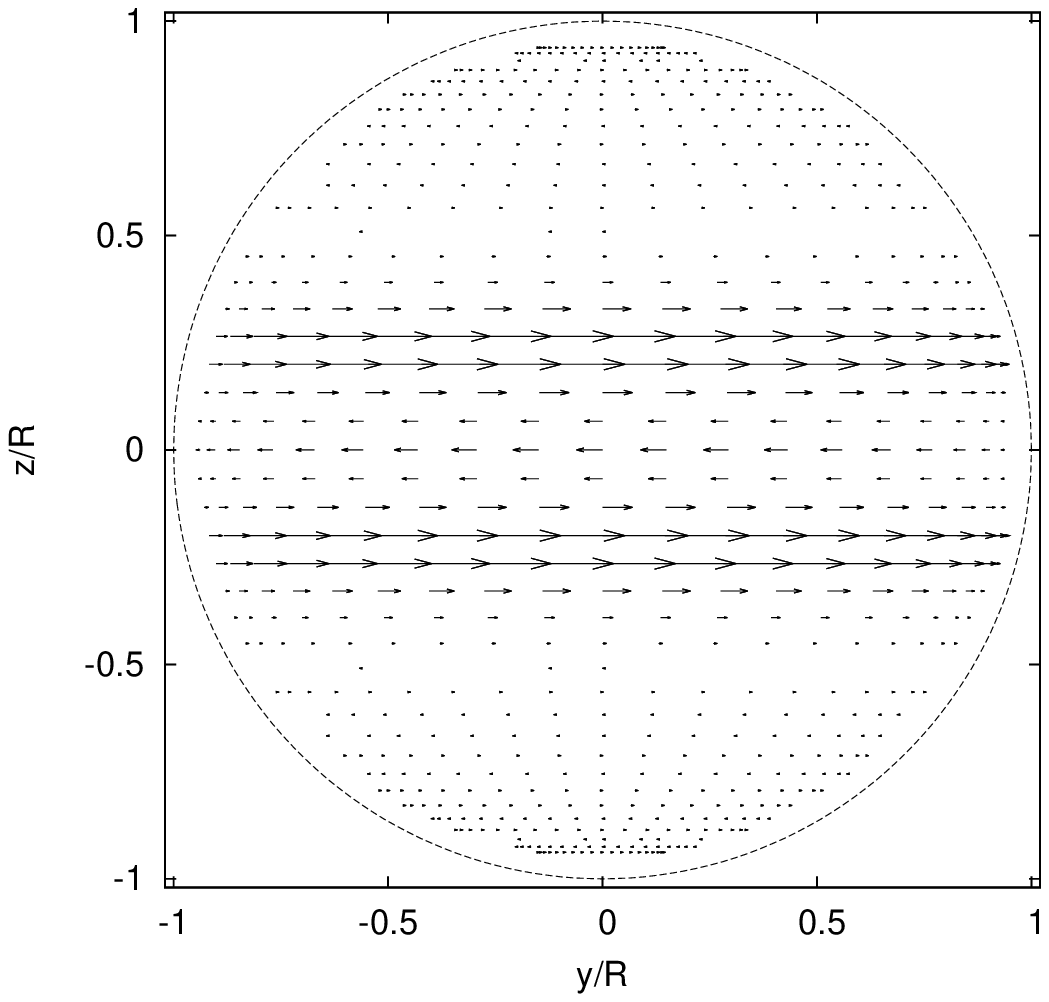}}
\resizebox{0.33\columnwidth}{!}{
\includegraphics{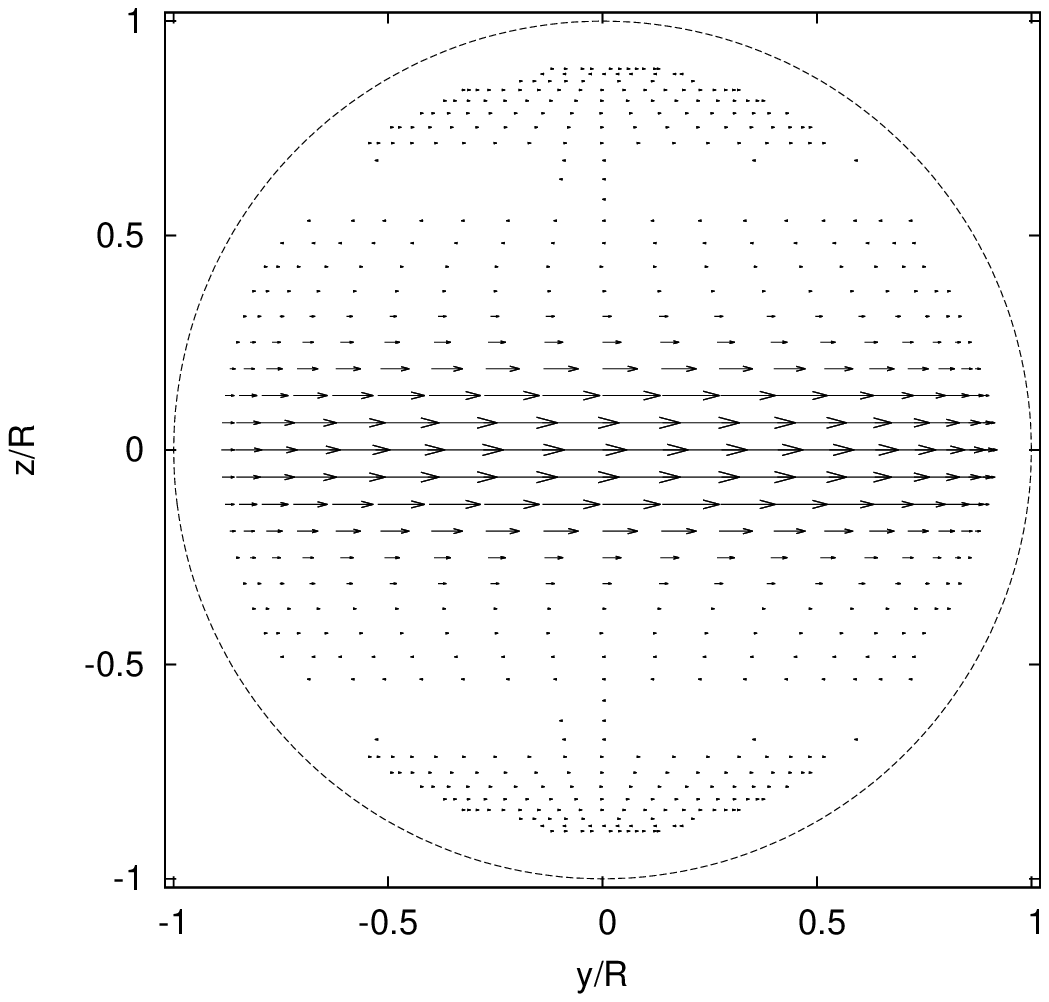}}
\end{center}
\caption{Same as Figure \ref{fig:m15progreso} but for $\bar\omega_{\rm tide}=-0.3812296926$ in resonance with the retrograde $l=-m=2$ $g_7$-mode, where the ratio of the maximum velocity $v_{\rm max}^{(2)}(x)$ to that on the surface of
$x=0.99$ is 0.12 for $x=0.95$ and 0.04 for $x=0.90$, respectively.
}
\label{fig:g7reso_retro}
\end{figure}

\begin{figure}
\begin{center}
\resizebox{0.33\columnwidth}{!}{
\includegraphics{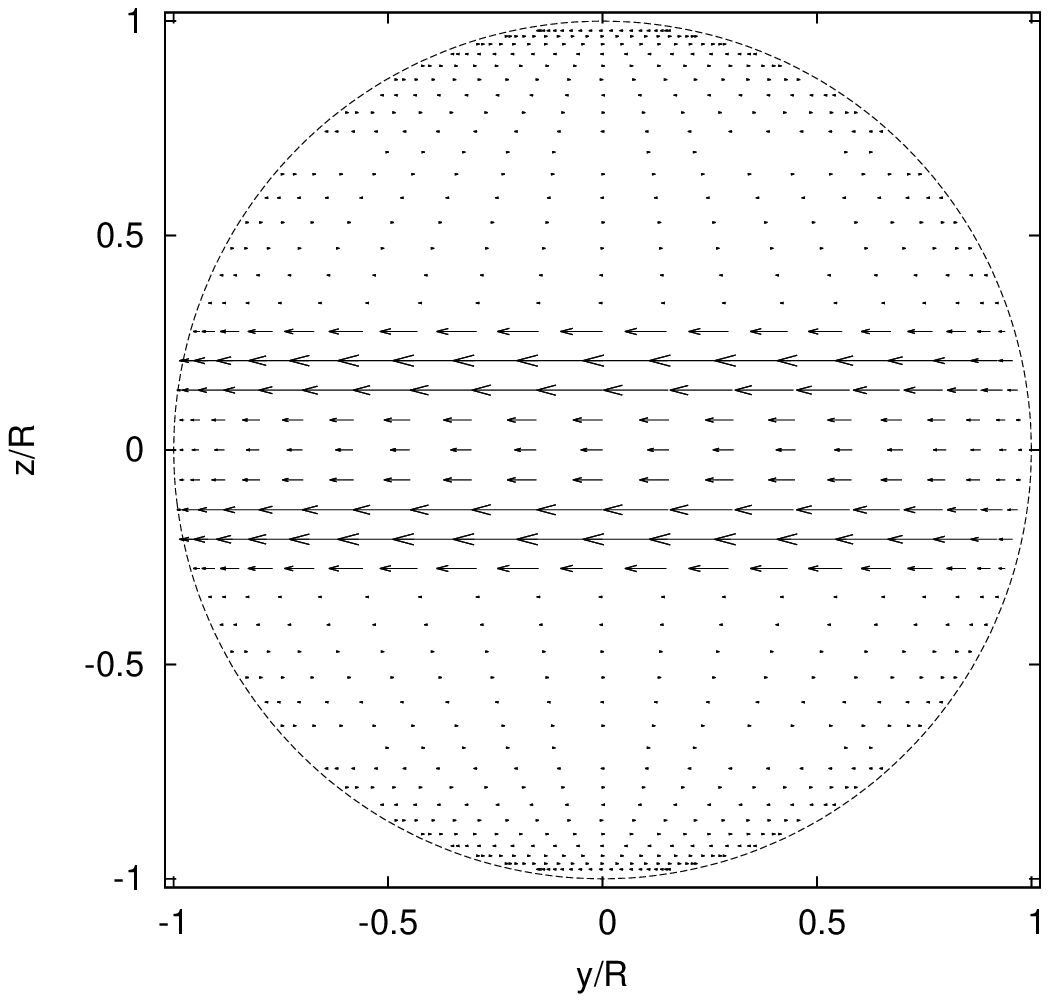}}
\resizebox{0.33\columnwidth}{!}{
\includegraphics{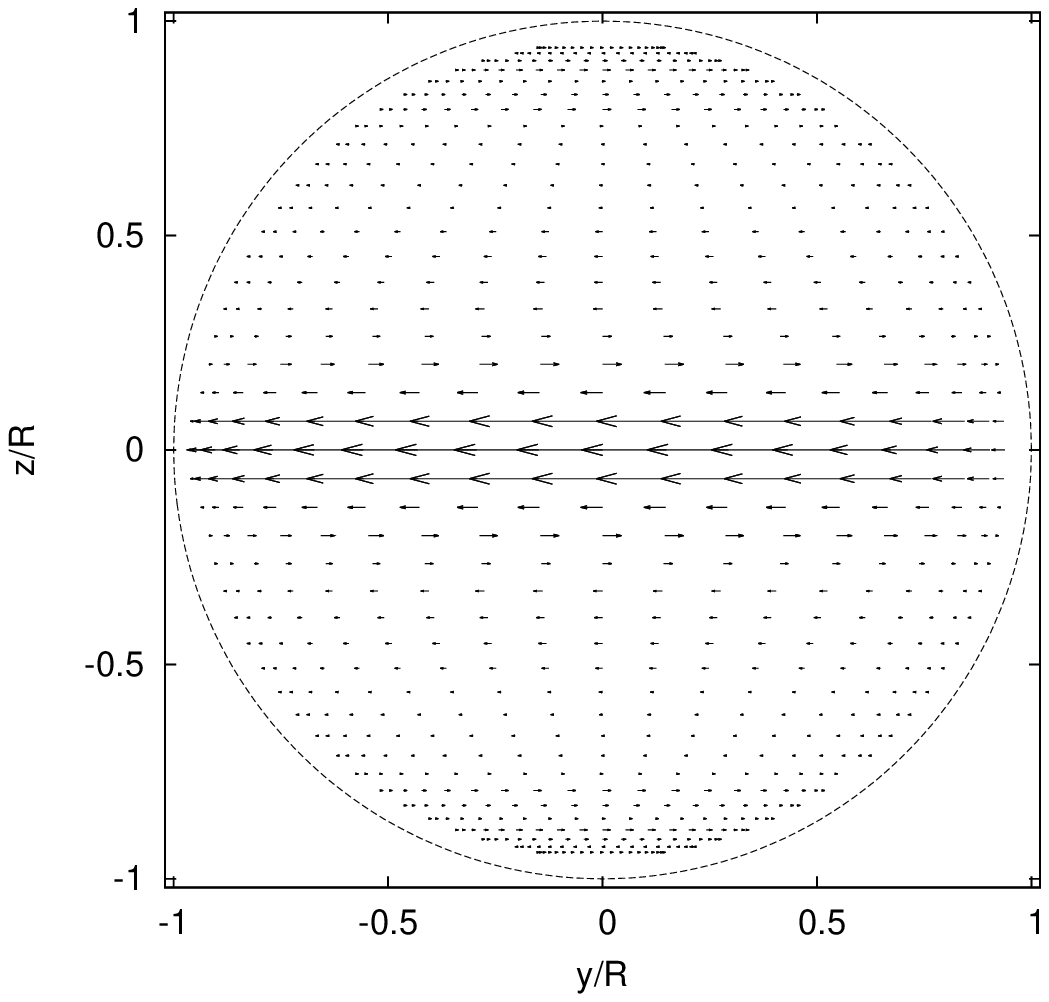}}
\resizebox{0.33\columnwidth}{!}{
\includegraphics{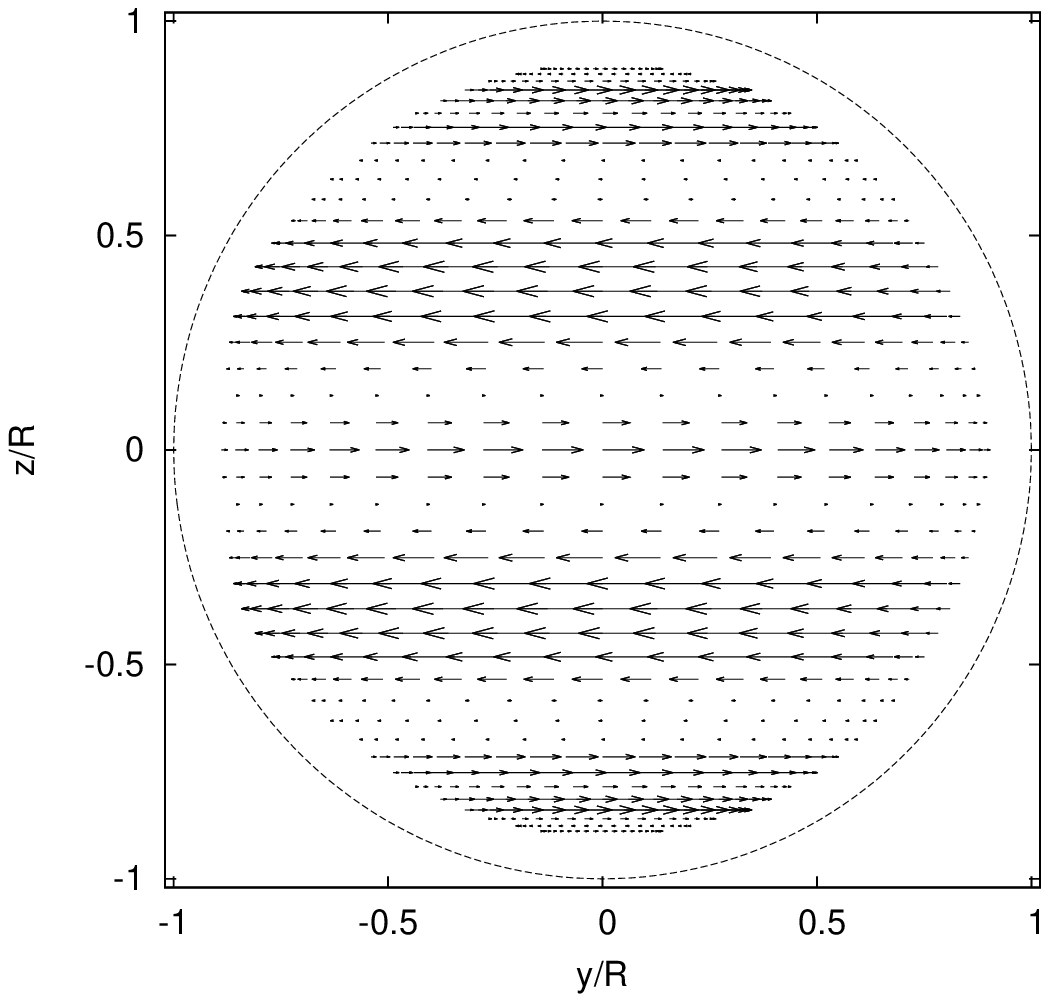}}
\end{center}
\caption{Same as Figure \ref{fig:m15progreso} but for $\bar\omega_{\rm tide}=-0.36$ in off-resonance 
with low frequency modes, where the ratio of the maximum velocity $v_{\rm max}^{(2)}(x)$ to that on the surface of
$x=0.99$ is 0.067 for $x=0.95$ and 0.012 for $x=0.90$, respectively.
}
\label{fig:g_offreso_retro}
\end{figure}

\begin{figure}
\begin{center}
\resizebox{0.45\columnwidth}{!}{
\includegraphics{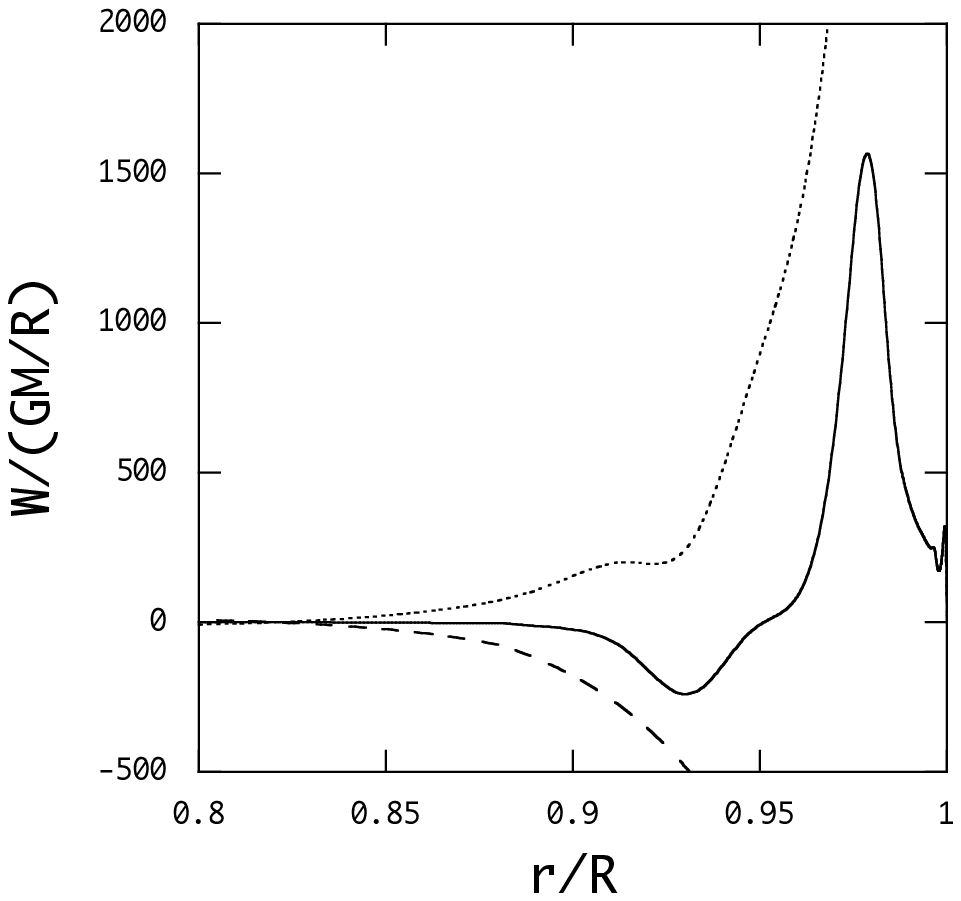}}
\resizebox{0.45\columnwidth}{!}{
\includegraphics{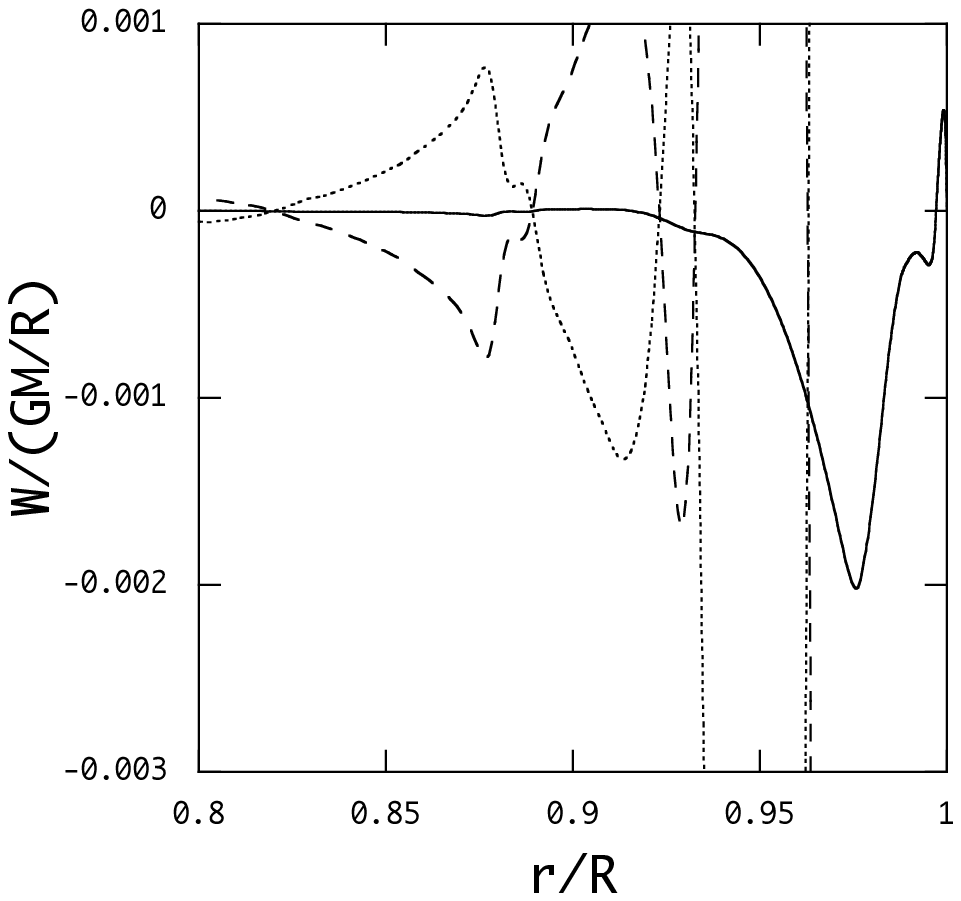}}
\end{center}
\caption{Function ${\cal W}$ (solid line) versus $x=r/R$ for the tidal responses at $\bar\omega_{\rm tide}=-0.3812296926$ (left panel), and
at $\bar\omega_{\rm tide}=-0.36$ (right panel) for the $15M_\odot$ ZAMS model, where $\bar\Omega=0.1$ and $f_0=1$ assumed.
The dotted and dashed lines indicate the first and second terms on the right-hand-side of equation (\ref{eq:defw}), respectively.
}
\label{fig:wfunctions_retro}
\end{figure}

Figure \ref{fig:vmax} plots $v_{\rm max}^{(2)}/(R\sigma_0)$ at $x=0.99$ as a function of the forcing frequency $\omega_{\rm tide}$ for the $15M_\odot$ model for $\bar\gamma=-10^{-8}$ and $f_0=1$.
The velocity $v_{\rm max}^{(2)}/(R\sigma_0)$ makes peaks at resonance with low frequency modes and can be as large as $10^{12}$
for low radial order $g$-modes, 
and the height of the peaks decreases as the radial order of $g$-modes increases.
We also note that $v_{\rm max}^{(2)}/(R\sigma_0)$ in off-resonance stays around $\sim 10^6$.
Since $v_{\rm max}^{(2)}/(R\sigma_0)\propto f_0^2\propto q^2(a_{\rm orb}/R)^{-6}$, if we assume $q\sim 0.1$, $v_{\rm max}^{(2)}/(R\sigma_0)$ 
at $\bar\omega_{\rm tide}$ in resonance with low radial order $g$-modes will be $\sim 10^4$ for $a_{\rm orb}/R\sim 10$
and $\sim 10^{-2}$ for $a_{\rm orb}/R\sim 10^2$.

\begin{figure}
\begin{center}
\resizebox{0.45\columnwidth}{!}{
\includegraphics{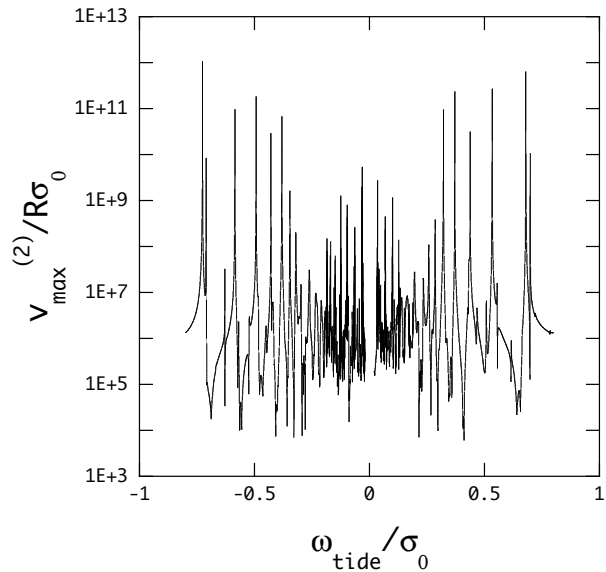}}
\end{center}
\caption{$v_{\rm max}^{(2)}/R\sigma_0$ at $x=0.99$ as a function of the forcing frequency $\bar\omega_{\rm tide}$ for $\bar\Omega=0.1$, $f_0=1$, and $\bar\gamma=-10^{-8}$.
}
\label{fig:vmax}
\end{figure}

\section{conclusion}

In this paper, we computed tidally driven axisymmetric mean flows in a slowly and uniformly rotating massive main sequence star in a binary system, assuming that the tidal potential due to the companion star
is a small perturbation to the primary star and the mean flows excited in the primary are of second order of the perturbation amplitudes.
Here, we ignored equilibrium structure deformation caused by rotation and tidal force so that the equilibrium structure can be treated as being spherical symmetric.
To compute the mean flows, we made a simplifying assumption that the time derivatives $\partial\pmb{v}^{(2)}/\partial t$
can be replaced by $\gamma\pmb{v}^{(2)}$ where $\gamma$ is a constant parameter regarded as the growth (or decay) rate of the second order perturbations.
We find that the $\phi$-component of the velocity fields $\pmb{v}^{(2)}$ is the dominant one and 
that the amplitudes tend to be confined in the equatorial regions in the surface layers and decrease as we go into the deep interior where non-adiabatic effects become insignificant.
We find the velocities $\pmb{v}^{(2)}$ in the deep interior are prograde (retrograde) for the prograde (retrograde) forcing $\omega_{\rm tide}$, which may be consistent with the picture that dissipation in the deep interior associated with tidal responses cause synchronization between orbital motion and stellar rotation
in binary systems.
We also discussed the relation between the term $\partial v_\phi^{(2)}/\partial t$ averaged over the colatitude $\theta$ and the function ${\cal W}(r)$,
assuming the averaged $\partial v_\phi^{(2)}/\partial t$ is the dominant term in the angular momentum conservation equation.

The velocities $v_\phi^{(2)}$ of tidally driven mean flows depend on both $r$ and $\theta$, which inevitably leads to differential rotation in the interior in the time scales of order of $\sim \gamma^{-1}$.
In this paper, we assumed that the star is uniformly rotating when computing tidal responses and that
the time dependence of tidally driven mean flows is given by $e^{\gamma t}$ and that of the responses by
$e^{\gamma t/2}$ to derive the governing equations for mean flows of second order.
Probably, this is not necessarily a good approximation for the problem when we consider binary evolution
in the time scales longer than $\sim \gamma^{-1}$, in which time scales equilibrium rotation laws would 
become substantially different from uniform rotation.
It is thus highly desirable to follow time development of mean flows as a result of 
interactions between the mean flows and tidal responses in differentially rotating stars.

We should be cautious about the results suggested by Figure \ref{fig:vmax}, where
we have computed $v_{\rm max}^{(2)}/(R\sigma_0)$ at $x=0.99$ as a function of $\bar\omega_{\rm tide}$
assuming that $\bar\gamma$ is a constant.
For example, however, if we assume that $\bar\gamma$ is given by $\bar\gamma_{\rm tide}\propto {\cal T}_2$, 
$\bar\gamma$ will make sharp resonance peaks as a function of $\bar\omega_{\rm tide}$.
The rapid increase in $|\bar\gamma|$ at peaks, on the other hand, will suppress the resonance peaks of $v_{\rm max}^{(2)}/(R\sigma_0)$
when $v_{\rm max}^{(2)}\bar\gamma\sim C$ holds where $C$ is a constant that does not depend on $\bar\omega_{\rm tide}$.
This suggests that the magnitudes of $v_{\rm max}^{(2)}$ will be only weakly dependent on $\bar\omega_{\rm tide}$
even near resonance although the flow patterns in resonance will be different from those in off-resonance.
If this is the case for $\bar\gamma=\bar\gamma_{\rm tide}$, we can estimate the magnitudes of $v_{\rm max}^{(2)}/R\sigma_0$ at $x=0.99$ using the numerical results
obtained for tidal resonance with low radial order $g$-modes.
As suggested in the last paragraph of the previous section, for the parameter values of $q\sim0.1$ and $a_{\rm orb}/R\sim 10$,
for example, 
we have $\bar\gamma_{\rm tide}\sim 10^{-8}$, for which the magnitude of $v_{\rm max}^{(2)}/R\sigma_0$ will be
of order of $\sim 10^{4}$.
This value is too large to be accepted.
Obviously we need more careful analyses concerning possible amplitudes of the mean flows driven by tidal responses.

In this paper, we assumed that the star is slowly rotating at $\bar\Omega=0.1$.
For slow rotation, low radial order $g$-modes are not necessarily significantly affected by
rotation, and there arise no significant differences in the mode properties between prograde and retrograde low radial order $g$-modes, although there appear on the retrograde side sequences of $r$-modes whose oscillation frequency in the co-rotating frame of the star is comparable to or less than $\Omega$.
For rapidly rotating stars, as suggested by Figure 2, the tidal responses will have properties 
qualitatively different from those in slowly rotating stars, even in the frequency ranges of low radial order $g$-modes.
The properties of tidal responses of a massive star also depend on the evolutional stages.
As the star evolves from the ZAMS stage, the frequency spectra of low frequency $g$-modes
will be denser and the amplitudes of $g$-modes tend to be confined into the deep interior.
The development of a $\mu$-gradient zone outside the convective core will make the frequency spectra more complicated.
Since low frequency $g$-modes can be trapped in the well-developed $\mu$-gradient zone, if the tidal forcing is 
in resonance with $g$-modes trapped in the $\mu$-zone, mean flows driven by the $g$-modes will have 
mixing effects on material there even if non-adiabatic effects are small in the deep interior.

Tidal responses and tidally driven mean flows of the star discussed in this paper have rather simple
properties since no low frequency modes of the model are pulsationally unstable.
Probably, this is not the case for slowly pulsating (SPB) stars, because
numerous low frequency $g$-modes and $r$-modes of the stars are destabilized by the opacity bump mechanism.
For these variable stars, there exists a strong excitation zone that surpasses damping contributions in the interior.
The sign of the tidal torque ${\cal T}_2$ may change as a function of $\omega_{\rm tide}$.
We expect that tidal mean flows driven in SPB stars will have different properties
from those in massive main sequence stars.


\begin{appendix}

\section{differential equations for tidal responses}

Substituting the series expansions (\ref{eq:xiexp_r}) to (\ref{eq:pexpansion}) into
the perturbed basic equations (\ref{eq:linearizedeom2}), (\ref{eq:eqcontinuity}), (\ref{eq:eqentropy}), and (\ref{eq:eos}), we obtain
a finite set of linear ordinary differential equations for the expansion coefficients (see, e.g., Lee \& Saio 1987).
If we use vector notation for the set of differential equations,
defining the dependent variables $\pmb{y}_j$, $\pmb{h}$, $\pmb{t}$, and $\pmb{\psi}$ as
\be
\pmb{y}_1=\left(S_l\right), \quad \pmb{y}_2=\left({p'_l\over\rho gr}\right), \quad \pmb{y}_3=\left({\delta L_{{\rm rad},l}\over L_{\rm rad}}\right), \quad 
\pmb{y}_4=\left({\delta s_l\over c_p}\right), \quad \pmb{h}=\left(H_l\right), \quad \pmb{t}=\left(T_{l'}\right), \quad \pmb{\psi}=\left(\Phi^\prime_{e,l}+\Psi_l\right), 
\ee
we write the set of linear ordinary differential equations for tidally forced non-adiabatic oscillations of rotating stars as
\be
r{\partial\pmb{y}_1\over \partial r}=
\left[\left({V\over\Gamma_1}-3\right)\pmbmt{I}+q\pmbmt{WO}\right]\pmb{y}_1
+\left({\pmbmt{W}\over c_1\bar\omega^2}-{V\over\Gamma_1}\pmbmt{I}\right)\pmb{Y}_{2}
+\alpha_T\pmb{y}_4+{V\over\Gamma_1}{\pmb{\psi}\over gr},
\label{eq:y111}
\ee
\begin{eqnarray}
r{\partial\pmb{Y}_2\over \partial r}=\left[\left(c_1\bar\omega^2+rA\right)\pmbmt{I}-4c_1\bar\Omega^2\pmbmt{G}\right]\pmb{y}_1
+\left[\left(1-U-rA\right)\pmbmt{I}-q\pmbmt{O}^T\pmbmt{W}\right]\pmb{Y}_2
+\alpha_T\pmb{y}_4+rA{\pmb{\psi}\over gr},
\label{eq:y222}
\end{eqnarray}
\begin{eqnarray}
r{\partial\pmb{y}_3\over\partial r}&=&\left(E_1\pmbmt{I}-
\beta\pmbmt{\Lambda}_0+E_0q\pmbmt{WO}\right)\pmb{y}_1
+\left(-E_1\pmbmt{I}-{\nabla_{ad}\over\nabla}\pmbmt{\Lambda}_0+E_0{\pmbmt{W}\over c_1\bar\omega^2}\right)\pmb{Y}_2-E_0\pmb{y}_3
\nonumber\\
&&+\left\{\left[\left(E_0-c_3\right)\alpha_T+c_3\epsilon_T-\rmi\omega c_2\right]\pmbmt{I}-{1\over\nabla V}\pmbmt{\Lambda}_0\right\}\pmb{y}_4+\left(E_1\pmbmt{I}+{\nabla_{ad}\over\nabla}\pmbmt{\Lambda}_0\right){\pmb{\psi}\over gr},
\label{eq:y333}
\end{eqnarray}
\begin{eqnarray}
{1\over\nabla V}r{\partial\pmb{y}_4\over\partial r}&=&\left\{\left[4\beta
+{\nabla_{ad}\over\nabla}\left(U-c_1\bar\omega^2\right)+E_2\right]\pmbmt{I}-\beta q\pmbmt{WO}+{\nabla_{ad}\over\nabla}4c_1\bar\Omega^2\pmbmt{G}\right\}\pmb{y}_1
-\left(E_2\pmbmt{I}+\beta{\pmbmt{W}\over c_1\bar\omega^2}-{\nabla_{ad}\over\nabla}q\pmbmt{O}^T\pmbmt{W}\right)\pmb{Y}_2
\nonumber\\
&&-\pmb{y}_3+\left(4-\kappa_T\right)\pmb{y}_4
+E_2{\pmb{\psi}\over gr},
\label{eq:y444}
\end{eqnarray}
where
\be
\pmb{Y}_2=\pmb{y}_2+{\pmb{\psi}\over gr},
\ee
and
\be
q={2\Omega\over\omega}, \quad \bar\omega={\omega\over\sigma_0}, \quad \bar\Omega={\Omega\over\sigma_0}, \quad
\sigma_0=\sqrt{GM\over R^3},
\ee
\be
V=-{d\ln p\over d\ln r}, \quad U={d\ln M_r\over d\ln r},
\ee
\be
\nabla={d\ln T\over d\ln p}, \quad \nabla_{ad}=\left({\partial\ln T\over\partial \ln p}\right)_{ad}, \quad \beta=1-{\nabla_{ad}\over\nabla},
\ee
\be
c_1={(r/R)^3\over M_r/M}, \quad c_2={4\pi r^3\rho Tc_p\over L_{\rm rad}}\sigma_0, \quad
c_3={4\pi r^3\rho\epsilon\over L_{\rm rad}},
\ee
\be
\epsilon_{ad}=\left({\partial\ln\epsilon\over\partial\ln p}\right)_{ad}, \quad
\epsilon_T=\left({\partial\ln\epsilon\over\partial\ln T}\right)_T, \quad
\kappa_{ad}=\left({\partial\ln\kappa\over\partial\ln p}\right)_{ad}, \quad
\kappa_T=\left({\partial\ln\kappa\over\partial\ln T}\right)_T,
\ee
\be
E_0={d\ln L_{\rm rad}\over d\ln r}, \quad 
E_1=\left(E_0-c_3\right){V\over \Gamma_1}-c_3\epsilon_{ad}V, \quad
E_2=\left(-4\nabla_{ad}+\kappa_{ad}\right)V+{\nabla_{ad}\over\nabla}\left(V+{d\ln\nabla_{ad}\over d\ln r}\right).
\ee
Note that the relations between the variables $(\pmb{h},\rmi\pmb{t})$ and $(\pmb{y}_1,\pmb{Y}_2)$ are given 
\be
\pmbmt{\Lambda}_0\pmb{h}
={\pmbmt{W}\over c_1\bar\omega^2}\pmb{Y}_2+q\pmbmt{WO}\pmb{y}_1,
\label{eq:lamh}
\ee
\be
2c_1\bar\omega\bar\Omega\left(m\pmb{h}+\rmi\pmb{t}\right)
=q\pmbmt{O}^T\pmbmt{W}\pmb{Y}_2+4c_1\bar\Omega^2\pmbmt{G}\pmb{y}_1,
\label{eq:mhpit}
\ee
where 
\be
\pmbmt{W}=\pmbmt{\Lambda}_0\left(\pmbmt{L}_0-\pmbmt{M}_1\pmbmt{L}_1^{-1}\pmbmt{M}_0\right)^{-1}, \quad
\pmbmt{O}=m\pmbmt{\Lambda}_0^{-1}-\pmbmt{M}_1\pmbmt{L}_1^{-1}\pmbmt{K}, \quad
\pmbmt{G}=\pmbmt{O}^T\pmbmt{WO}-\pmbmt{C}_0\pmbmt{L}_1^{-1}\pmbmt{K},
\ee
and $\pmbmt{O}^T$ is the transpose matrix of $\pmbmt{O}$, $\pmbmt{I}$ is the unit matrix.
The non-zero elements of the matrices $\pmbmt{\Lambda}_0$, $\pmbmt{\Lambda}_1$, $\pmbmt{L}_0$, $\pmbmt{L}_1$,
$\pmbmt{M}_0$, $\pmbmt{M}_1$, $\pmbmt{K}$, $\pmbmt{C}_0$ for even modes
are
\be
\left(\pmbmt{\Lambda}_0\right)_{j,j}=l_j(l_j+1),\quad \left(\pmbmt{\Lambda}_1\right)_{j,j}=l'_j(l'_j+1),\quad
\left(\pmbmt{L}_0\right)_{j,j}=1-{mq\over l_j(l_j+1)}, \quad \left(\pmbmt{L}_1\right)_{j,j}=1-{mq\over l'_j(l'_j+1)},
\ee
\be
\left(\pmbmt{M}_0\right)_{j,j}=q{l_j\over l_j+1}J^m_{l_j+1}, \quad \left(\pmbmt{M}_0\right)_{j,j+1}=q{l_j+3\over l_j+2}J^m_{l_j+2}, \quad
\left(\pmbmt{M}_1\right)_{j,j}=q{l_j+2\over l_j+1}J^m_{l_j+1}, \quad \left(\pmbmt{M}_1\right)_{j+1,j}=q{l_j+1\over l_j+2}J^m_{l_j+2},
\ee
\be
\left(\pmbmt{K}\right)_{j,j}={J^m_{l_j+1}\over l_j+1}, \quad \left(\pmbmt{K}\right)_{j,j+1}=-{J^m_{l_j+2}\over l_j+2}, \quad
\left(\pmbmt{C}_0\right)_{j,j}=-\left(l_j+2\right)J^m_{l_j+1}, \quad \left(\pmbmt{C}_0\right)_{j+1,j}=\left(l_j+1\right)J^m_{l_j+2},
\ee
and for odd modes
\be
\left(\pmbmt{\Lambda}_0\right)_{j,j}=l_j(l_j+1),\quad \left(\pmbmt{\Lambda}_1\right)_{j,j}=l'_j(l'_j+1),\quad
\left(\pmbmt{L}_0\right)_{j,j}=1-{mq\over l_j(l_j+1)}, \quad \left(\pmbmt{L}_1\right)_{j,j}=1-{mq\over l'_j(l'_j+1)},
\ee
\be
\left(\pmbmt{M}_0\right)_{j,j}=q{l'_j+2\over l'_j+1}J^m_{l'_j+1}, \quad \left(\pmbmt{M}_0\right)_{j+1,j}=q{l'_j+1\over l'_j+2}J^m_{l'_j+2}, \quad
\left(\pmbmt{M}_1\right)_{j,j}=q{l'_j\over l'_j+1}J^m_{l'_j+1}, \quad \left(\pmbmt{M}_1\right)_{j,j+1}=q{l'_j+3\over l'_j+2}J^m_{l'_j+2},
\ee
\be
\left(\pmbmt{K}\right)_{j,j}=-{J^m_{l'_j+1}\over l'_j+1}, \quad \left(\pmbmt{K}\right)_{j+1,j}={J^m_{l'_j+2}\over l_j+2},
\left(\pmbmt{C}_0\right)_{j,j}=l'_jJ^m_{l'_j+1}, \quad \left(\pmbmt{C}_0\right)_{j,j+1}=-\left(l'_j+3\right)J^m_{l'_j+2}\ee
where $l_j=2(j-1)+|m|$ and $l_j'=l_j+1$ for even modes and $l_j=2j-1+|m|$ and $l_j'=l_j-1$ for odd modes, and
\be
J^m_j=\sqrt{l^2-m^2\over 4l^2-1}
\ee
for $l\ge|m|$, and $J^m_l=0$ otherwise.

We note that the terms proportional to $\pmb{\psi}$ are inhomogeneous terms of the set of 
linear differential equations, and if we drop these inhomogeneous terms the set of linear ordinary differential
equations reduce to those for free oscillations of stars (Lee \& Saio 1987).
The oscillation frequency $\omega$ should be regarded as the tidal forcing frequency $\omega_{\rm tide}$ for tidal responses.

To integrate the set of linear ordinary differential equations, we employ a Henyey type method of integration.
For free oscillations of stars, for example, we formally write the set of linear differential equations as
\be
{d\pmb{Y}\over dx}=\pmbmt{C}(x,\omega)\pmb{Y}, 
\label{eq:dydx}
\ee
where $x=\ln r$, 
\be
\pmb{Y}=\left(\matrix{\pmb{y}_1\cr\pmb{y}_2\cr\pmb{y}_3\cr\pmb{y}_4\cr}\right),
\ee
and $\pmbmt{C}$ is the coefficient matrix.
The differential equation (\ref{eq:dydx}) may reduce to a set of difference equations given by
\be
{\pmb{Y}^{n+1}-\pmb{Y}^n\over \Delta x^{n+1/2}}=\alpha\pmbmt{C}^{n+1}\pmb{Y}^{n+1}+(1-\alpha)\pmbmt{C}^{n}\pmb{Y}^{n},
\quad \pmbmt{C}^{n}=\pmbmt{C}(x^n,\omega), \quad \Delta x^{n+1/2}=x^{n+1}-x^n,
\label{eq:differenceeq}
\ee
where $n$ is the mesh number of the background model, running from $n=1$ (the center) to $n=N$ (the surface of
the model), and we usually assume $\alpha=1/2$.
Equations (\ref{eq:differenceeq}) give recurrence equations
\be
\pmbmt{S}^n\pmb{Y}^{n+1}+\pmbmt{T}^n\pmb{Y}^n=\pmb{d}^n,
\label{eq:recc}
\ee
where
\be
\pmbmt{S}^n=\pmbmt{I}-\Delta x^{n+1/2}\alpha\pmb{C}^{n+1}, \quad \pmbmt{T}^n=-\pmbmt{I}-\Delta x^{n+1/2}(1-\alpha)\pmbmt{C}^n, \quad \pmb{d}^n=0.
\ee
The inner and outer boundary conditions and the amplitude normalization may be
written as
\be
\pmbmt{B}_{\rm in}\pmb{Y}^1=0, \quad \pmbmt{B}_{\rm out}\pmb{Y}^N=0, \quad S_{l_1}^N=1,
\label{eq:bbc}
\ee
where $\pmbmt{B}_{\rm in}$ and $\pmbmt{B}_{\rm out}$ are the coefficient matrices defining the boundary conditions.
Using Newton-Raphson method, we look for $\omega$ such that the functions $\pmb{Y}^n$ satisfy all the recurrence relations (\ref{eq:recc}), the boundary conditions, and amplitude normalization (\ref{eq:bbc}).
The background models we use in this paper have more than 2000 mesh points in the radial direction, which makes it possible for us to get
accurate eigenmodes even when the modes have radial nodes of the eigenfunctions as many as $\sim 100$.
Note that for tidally forced oscillations the vectors $\pmb{d}^n$ become nonzero vectors because of the inhomogeneous terms due to $\pmb{\psi}$ and that we omit the normalization $S_{l_1}^N=1$ to calculate forced oscillations.

\end{appendix}



\bsp	
\label{lastpage}
\end{document}